\documentclass[final,5p,twocolumn]{elsarticle}

\usepackage[printonlyused]{acronym}
\acrodef{AVS}{acoustic vector-sensor}
\acrodef{ASR}{automatic speech recognition}
\acrodef{SNR}{signal-to-noise ratio}
\acrodef{LTI}{linear time invariant}
\acrodef{STFT}{short-time Fourier transform}
\acrodef{DRR}{direct-reverberation ratio}
\acrodef{RTF}{relative transfer function}
\acrodef{SDW-MWF}{speech distortion weighted multichannel Wiener filter}
\acrodef{MVDR}{minimum variance distortionless response}
\acrodef{MSE}{mean square error}
\acrodef{MWF}{multichannel Wiener filter}
\acrodef{HATS}{head and torso simulator}
\acrodef{STOI}{short-time objective intelligibility}
\acrodef{MPDR}{minimum power distortionless response}
\acrodef{MWF}{multichannel Wiener filter}
\acrodef{SWF}{single-channel Wiener filter}
\acrodef{SDW-MWF}{Speech distortion weighted multichannel Wiener filter}
\acrodef{DOA}{direction of arrival}
\acrodef{MEMS}{microelectromechanical systems}
\acrodef{RA}{reduced array}

\def\bh{{\mathbf{h}}}
\def\bx{{\mathbf{x}}}
\def\bg{{\mathbf{g}}}
\def\be{{\mathbf{e}}}
\def\ba{{\mathbf{a}}}
\def\bv{{\mathbf{v}}}
\def\bc{{\mathbf{c}}}
\def\bb{{\mathbf{b}}}
\def\bw{{\mathbf{w}}}
\def\hr{{\widetilde{\bh}_{\mathrm{d}}}}
\def\sr{{\widetilde{s}}}
\def\bPhi{{\mathbf{\Phi}}}
\def\bPhihat{{\widehat{\mathbf{\Phi}}}}
\def\H{{\mathrm{H}}}
\def\T{{\mathrm{T}}}
\def\sest{{\widehat{s}}}
\def\WF{W}
\def\rd{{\mathrm{d}}} 
\def\bu{{\mathbf{u}}}
\def\bq{{\mathbf{q}}}

\newcommand{\ed}[1]{#1}
\newcommand{\argmin}{\operatornamewithlimits{argmin}}

\newcommand{\bed}[1]{#1}
\newcommand{\red}[1]{#1}
\newcommand{\bdot}{{\boldsymbol{\cdot}}}

\usepackage{amssymb}
\usepackage{amsmath}
\usepackage{enumitem}
\usepackage{subfigure}
\usepackage[update, prepend]{epstopdf}
\usepackage{color}
\usepackage{placeins}
\usepackage{balance}
\usepackage[lined, algonl, ruled]{algorithm2e}
\usepackage[hyphens]{url}

\journal{arXiv}
\usepackage[table]{xcolor}
\definecolor{lightgray}{gray}{0.9}

\usepackage{hyperref}
\hypersetup{
	bookmarks=true,
	bookmarksopen = false,
	bookmarksnumbered = true,
	colorlinks=true,
	hidelinks,
	pdfstartview = {FitH},
	pdftitle = {Near-field signal acquisition for smartglasses using two acoustic vector-sensors},
	pdfauthor = {Dovid Y.~Levin, Emanu\"{e}l A.~P. Habets, and Sharon Gannot}
	}

\begin{document}

\begin{frontmatter}
		
		
		
\title{\bf{Near-field signal acquisition for smartglasses using two acoustic vector-sensors}
	\tnoteref{t1}}
	\tnotetext[t1]{{\scriptsize$\copyright$} 2016.  This manuscript version is made available under the CC-BY-NC-ND 4.0 license \url{http://creativecommons.org/licenses/by-nc-nd/4.0/}~;
		published version from \emph{Elsevier Speech Communication} at
		\url{dx.doi.org/10.1016/j.specom.2016.07.002}~.}
\author[biu]{Dovid Y.~Levin\corref{CA}}
\ead{david.levin@live.biu.ac.il}
\author[ial]{Emanu\"{e}l A.~P. Habets}
\ead{emanuel.habets@audiolabs-erlangen.de}
\author[biu]{Sharon Gannot}
\ead{Sharon.Gannot@biu.ac.il}
		
\address[biu]{Bar-Ilan University, Faculty of Engineering, Building 1103, Ramat-Gan, 5290002, Israel}
\address[ial]{International Audio Laboratories Erlangen$^\dag$,
Am Wolfsmantel 33, 91058 Erlangen, Germany \newline
{\hspace{0mm} $^\dag$\scriptsize{A joint institution of the Friedrich-Alexander-University Erlangen-N\"{u}rnberg (FAU) and Fraunhofer IIS, Germany.}}}

\cortext[CA]{Corresponding author}
		
\begin{abstract}
Smartglasses, in addition to their visual-output
capabilities, often contain acoustic sensors for receiving the user's voice.  However, operation in noisy environments may lead to significant degradation of the received signal. To address this issue, we propose employing an acoustic sensor array  which is mounted on the eyeglasses frames.  The signals from the array are processed by an algorithm with the purpose of acquiring the desired near-field speech signal produced by the wearer while suppressing noise signals originating from the environment.  The array is comprised of two \acp{AVS} which are located at the fore of the glasses' temples.  Each \ac{AVS} consists of four collocated subsensors: one pressure sensor (with an omnidirectional response) and three particle-velocity sensors (with dipole responses) oriented in mutually orthogonal directions.  The array configuration is designed to boost the input power of the desired signal, and to ensure that the characteristics of the noise at the different channels are sufficiently diverse (lending towards more effective noise suppression).  Since changes in the array's position correspond to the desired speaker's movement, the relative source-receiver position remains unchanged; hence, the need to track fluctuations of the steering vector is avoided.  Conversely, the spatial statistics of the noise are subject to rapid and abrupt changes due to sudden movement and rotation of the user's head. Consequently, the algorithm must be capable of rapid adaptation toward such changes.
We propose an algorithm which incorporates detection of the desired speech in the  time-frequency domain, and employs this information to adaptively update estimates of the noise statistics.  
The speech detection plays a key role in ensuring the quality of the output signal.
We conduct controlled measurements of the array in noisy scenarios.  The proposed algorithm preforms favorably with respect to conventional algorithms. 
\end{abstract}
		
\begin{keyword}
beamforming \sep acoustic vector-sensors \sep smartglasses \sep adaptive signal processing

\PACS 43.60.Fg \sep 43.60.Mn \sep 43.60.Hj
			
			
			
\end{keyword}
		
\end{frontmatter}

	
\section{Introduction}
Recent years have witnessed an increased interest in \emph{wearable computers} \cite{Randell05, Barfield16}.  These devices consist of miniature computers worn by users which can perform certain tasks; the devices may incorporate various sensors and feature networking capabilities.  For example, a \emph{smartwatch} may be used to display email messages, aid in navigation, and monitor the user's heart rate (in addition to functioning as a timepiece).

One specific type of wearable computer which has garnered much attention is the \emph{smartglasses} 
--- a device which displays computer generated information 
supplementing the user's visual field. A number of companies have been conducting research and development towards smartglasses 
intended for consumer usage \cite{Cass15} (e.g., Google Glass \cite{Ackerman13} and Microsoft HoloLens).
In addition to their visual-output capabilities, smartglasses 
may incorporate acoustic sensors.
These sensors are used for hands-free mobile telephony applications, and for 
applications using a voice-control interface
to convey commands and information to the device.

The performance of both of these applications suffers when operating in a noisy environment: in telephony, noise degrades the quality of the speech signal transmitted to the other party; similarly, the accuracy of \ac{ASR} systems is reduced when the desired speech is corrupted by noise.  A review of one prominent smartglasses prototype delineated these two issues as requiring improvement \cite{Sung14}.

To deal with these issues, we propose a system for the acquisition of the desired near-field speech in a noisy environment. The system is based on an acoustic array embedded in eyeglasses frames worn by the desired speaker.  The multiple signals received by the array contain both desired speech as well as undesired components.  These signals are processed by an adaptive beamforming algorithm to produce a single output signal with the aim of retaining the desired speech with little distortion while suppressing undesired components.

The scenario of a glasses mounted array presents some challenging features which are not encountered in typical speech processing.  Glasses frames constitute a spatially compact platform, with little room to spread the sensors out.  Typically, when sensors are closely spaced the statistical qualities of the noise at each sensor are highly 
correlated presenting difficulties in robust noise suppression \cite{Bitzer01}.  Hence, special
care must be taken in the design of the array.

The proposed array consists of two \acp{AVS} located, respectively, at the fore of the glasses' right and left temples.  In contrast to conventional sensors that measure only the pressure component of a sound field (which is a scalar quantity), an \ac{AVS} measures both the pressure and particle-velocity components.  An \ac{AVS} consists of four subsensors with different spatial responses: one omnidirectional sensor (corresponding to pressure) and three orthogonally oriented dipole sensors (corresponding to the components of the particle-velocity vector).  Hence, the array contains a total of eight channels (four from each \ac{AVS}). Since each subsensor possesses a markedly different spatial response, the statistical properties of the noise at the different subsensors are diverse. Consequently, robust beamforming is possible in spite of the limited spatial aperture. Another advantage afforded by the use of \acp{AVS} is that the dipole sensors amplify near-field signals more so than conventional omnidirectional sensors.  
Due to these sensors,
the desired speech signal (which is in the near-field due to the proximity to the sensors) undergoes a relative gain 
and is amplified with respect to the noise.  The \emph{relative gain} is 
explained and quantified in Sec.~\ref{sec:array}.
The interested reader is referred to the Appendix for further information 
	on \acp{AVS}.

		
The configuration in which the array is mounted on the speaker's glasses differs from the typical scenario in which a microphone array is situated in the environment of the user.
The glasses configuration possesses particular properties which lead to a number of benefits with respect to processing: (i) The close proximity of the desired source to the sensors leads to high \ac{SNR} which is favorable. (ii) For similar reasons, the reverberation of the desired speech is negligible with respect to its direct component, rendering dereverberation a nonissue. (iii) Any change in the location of the desired source brings about a corresponding movement of the array which is mounted thereon.  Consequently, the relative source-sensors configuration is essentially constant, precluding the need for tracking changes of the desired speaker's position.
	
Conversely, the glasses-mounted configuration presents a specific challenge.  The relative positions of the undesired acoustic sources with respect to the sensor array are liable to change rapidly.  For instance, when the user rotates his/her head the relative position of the array to external sound sources undergoes significant and abrupt changes.  This necessitates that the signal processing stage be capable of swift adaptation.

The proposed algorithm is based on \ac{MVDR} beamforming which is designed to minimize the residual noise variance under the constraint of maintaining a distortionless desired signal.  This type of beamforming was proposed by Capon \cite{Capon69} in the context of spatial spectrum analysis of seismic arrays.   Frost \cite{Frost72} employed this idea in the field of speech processing using a time-domain representation of the signals.  Later, Gannot \emph{et al.}~\cite{Gannot01} recast the \ac{MVDR} beamformer in the time-frequency domain.  In the current work, we adopt the time-frequency formulation.

In the proposed algorithm, the noise covariance matrix is adaptively estimated on an ongoing basis from the received signals.  
Since the received signals contain both desired and undesired components, 
the
covariance matrix obtained from a naive
implementation would contain significant
contributions of energy from the desired speech.  This is detrimental to
the performance of the processing. 	
To prevent desired speech from contaminating the noise covariance estimation, a speech detection component is employed.  Time-frequency bins which are deemed likely to contain desired speech are not used for estimating the noise covariance.

To further reduce noise,
the output of the \ac{MVDR} stage undergoes post-processing  
by a \ac{SWF}.
It has been shown \cite{Simmer01} that
application of \ac{MVDR} beamforming followed by a \ac{SWF} 
is optimal in the sense of minimizing the \ac{MSE}
[since it is equivalent to the \ac{MWF}].
	
The paper is structured as follows: Sec.~\ref{sec:array} describes the motivation guiding our specific array design.  In Sec.~\ref{sec:notation}, we introduce the notation used to describe the scenario in which the array operates and then present the problem formulation. Sec.~\ref{sec:algorithm} presents the proposed algorithm and how its various component interrelate.  Sec.~\ref{sec:evaluation} evaluates the performance of the proposed algorithm, and Sec.~\ref{sec:conclusion} concludes with a brief summary. 

	
\section{Motivation for array design}\label{sec:array}
In this section, we discuss the considerations which lead to our choices for the placement of the sensors and the types of sensors used.

\begin{figure}\label{Fig:frames}
\centering
\includegraphics[width=0.7\linewidth]{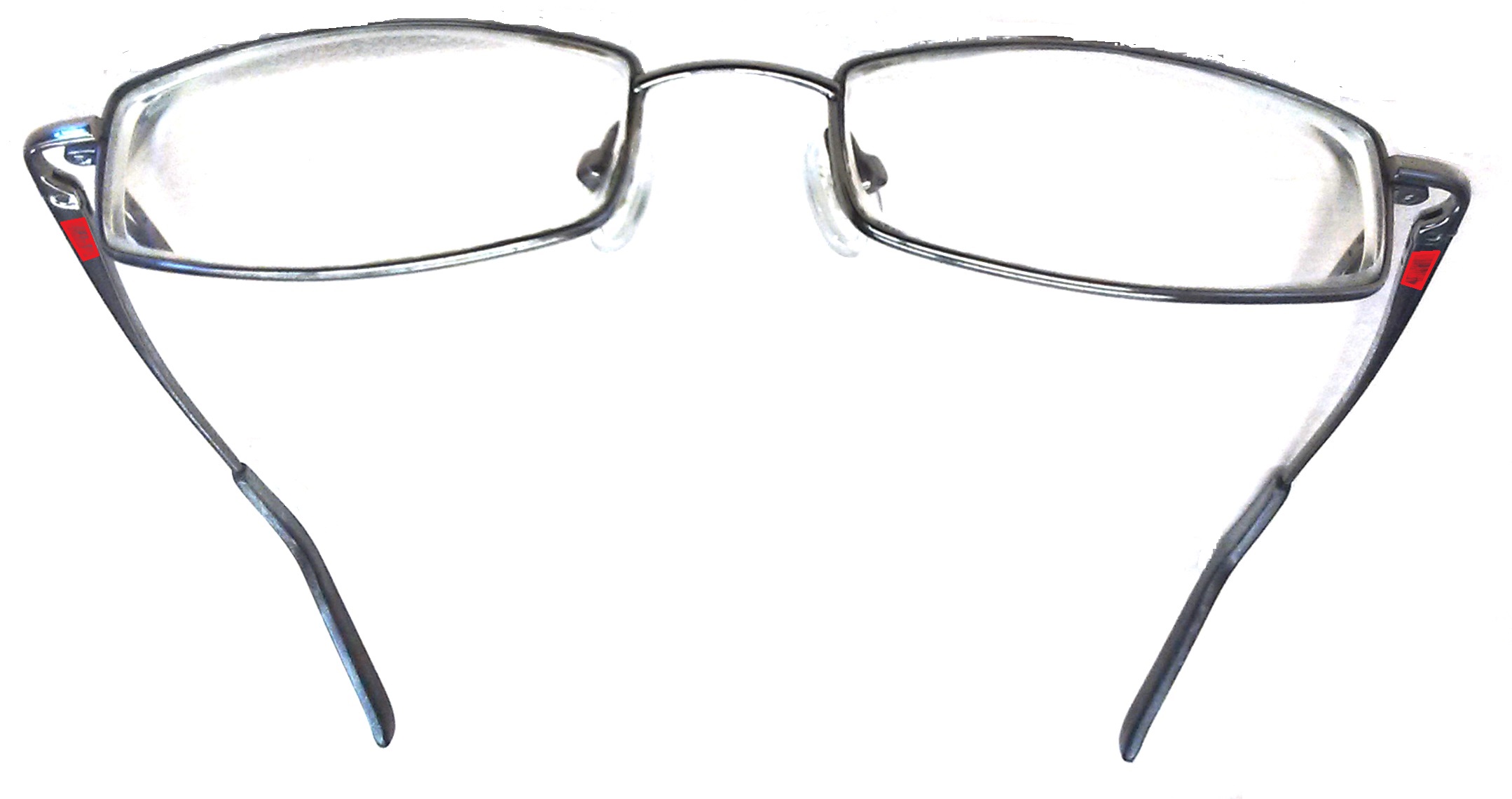}
\caption{The proposed sensor locations are indicated in red. \label{fig:frames2}}
\end{figure}

An \ac{AVS} is located at the fore of each of the glasses' temples (see Fig.~\ref{fig:frames2}).  The reason for selecting this location is that there is a direct ``line of sight'' path from the speaker's mouth to the sensors.  For other locations on the frames, such as the temples' rear sections or the areas above the lenses, the direct path is obstructed by human anatomy or the physical structure of the glasses.  The areas underneath the lenses were also considered as they \emph{do} have an unobstructed line to the mouth; however, embedding a microphone array at this locale was deemed to render the resulting frame structure too cumbersome.

Choosing an \ac{AVS} based array, rather than using conventional sensors, leads to several advantages.  Firstly, the inherent directional properties of an \ac{AVS} lend to the distinction between the desired source and sound arriving from other directions.  In contrast, a linear arrangement of conventional omnidirectional sensors along a temple of the glasses frame would exhibit a degree of directional ambiguity -- it is known that the response of such linear arrays maintains a conical symmetry \cite{VanTrees02}.
Secondly, an \ac{AVS} preforms well with a compact spatial configuration, whereas conventional arrays suffer from low robustness when array elements are closely spaced \cite{Bitzer01}.  Although this problem could be alleviated by allowing larger spacing between elements, this would necessitate placing sensors at the rear of the temple with no direct path to the source.  Thirdly, the near-field frequency response of dipoles amplifies lower frequencies. This effect, which results from the near-field acoustic impedance, tends to increase the \ac{SNR} since noise originating in the far-field does not undergo this amplification.  

To illustrate this last point, we consider the sensors' frequency responses to an ideal spherical wave\footnote{The actual propagation is presumably more complicated than an ideal spherical model and is difficult to model precisely.  For instance, the human mouth does not radiate sound uniformly in all directions. Furthermore, the structure of the human face may lead to some
diffraction
and reflection.  Nevertheless, the ideal spherical model is useful as it depicts overall trends.}\cite{Pierce91}.
The response of the monopole sensors is proportional to $\frac{1}{r}$, where $r$ is the distance from the wave's origin (i.e., they have a flat frequency response).  The response of the dipole elements is proportional to $\frac{1}{r}\left(1 +\frac{c}{r}\frac{1}{j\omega} \right)$ where $c$ is the velocity of sound propagation and $\omega$ is the angular frequency. Consequently, the dipoles have a relative gain of $\left(1 +\frac{c}{r}\frac{1}{j\omega} \right)$ over an omnidirectional sensor.  This becomes particularly significant at short distances and lower frequencies where $r \ll \frac{c}{\omega}$.
\bed{Stated differently, when the distance is significantly shorter than the wavelength, dipole sensors exhibit noticeable gain.}
\section{Notation and problem formulation}\label{sec:notation}
This section presents the scenario in which the array operates and the notation used to describe it.  The problem formulation is then presented
using this
notation.  

Let us denote the clean source signal as $s[n]$ and the $z_1[n]\ldots z_P[n]$ as $P$ interference signals.  These signals propagate from their respective sources to the sensors and may also undergo reflections inducing reverberation.  These processes are modeled as \ac{LTI} systems represented by impulse-responses.  Let $h_m[n]$ denote the response of the $m$-th sensor to an impulse produced by the desired source and $g_{m,p}[n]$ denote the impulse response from the $p$-th undesired source to the $m$-th sensor.  Each of the $M=8$ sensors is also subject to ambient noise 
and internal sensor noise; these will be denoted $e_m[n]$.  The resulting signal $x_m[n]$ received by the $m$-th sensor consists of all the above components and can be written as   
\begin{equation}\label{eq:composition}
	x_m[n] = s[n] * h_{m}[n] + \left(\sum_{p=1}^P z_{p}[n] * g_{m,p}[n]\right)
	 + e_m[n]\,. 
\end{equation}
Concatenating the respective elements into column-vectors, \eqref{eq:composition} can be reformulated as 
\begin{equation}\label{eq:composition2}
	\bx[n] = s[n] * \bh[n] + \left(\sum_{p=1}^P z_{p}[n] * \bg_{p}[n]\right) 
	+ \be[n] \,.
\end{equation}
The impulse response $\bh[n]$ can be decomposed into direct arrival and reverberation components $\bh[n] =  \bh_{\rd}[n] + \bh_{\mathrm{r}}[n]$.  The received signals can be expressed as
\begin{equation} \label{eq:composition3}
	\bx[n] = s[n] * \bh_\rd[n] + \bv[n]\,,
\end{equation}
where $\bv[n]$ incorporates the undesired sound sources, ambient and sensor noise, and reverberation of the desired source.  The vector $\bv[n]$ and all its different
subcomponents are referred to generically as \emph{noise} in this paper.
Since the sensors are mounted in close proximity to the mouth of the desired speaker, it can be assumed that the direct component is dominant 
with respect to reverberation
(i.e., the \ac{DRR} is high).  
	
The received signals are transformed to the time-frequency domain via the \ac{STFT}:
\begin{equation}
	\bx[n] 
	{\mapsto} \bx(\ell, k) =
	\begin{bmatrix}
	    X_1(\ell, k) \\
	    X_2(\ell, k) \\
	    \vdots \\
		X_M(\ell, k) 
	\end{bmatrix}\,,
\end{equation}
where the subscript denotes the channel and the indexes $\ell$ and $k$ represent time and frequency indexes, respectively.  A convolution in the time domain can be aptly approximated as multiplication in the \ac{STFT} domain provided that the analysis window is sufficiently long vis-\`{a}-vis the length of the impulse response \cite{Gannot01}.  Since the direct component of the impulse response is highly localized in time, $\bh_\rd[n]$ satisfies this criterion. Consequently, \eqref{eq:composition3} can be approximated in the \ac{STFT} domain as
\begin{equation}
	\bx(\ell, k)\, = \,\bh_\rd(k) s(\ell, k) + \bv(\ell, k)\,.
\end{equation}

Often the transfer function $\bh_\rd(k)$ is not available;  therefore, it is convenient to use the \ac{RTF} representation,
\begin{equation}
	\bx(\ell, k) \, = \, \hr(k) \sr(\ell, k) + \bv(\ell, k)\,,
\end{equation}
where $\sr$ is typically the direct component of the 
clean source signal
received at the first channel (or some linear combination of the different channels), and $\hr(k)$ is the \ac{RTF} of this signal with respect to the sensors.
Expressed formally,
	\begin{equation}\label{eq:cvec}
	\sr(\ell, k) = \bc^\H(k) \bh_\rd(k) s(\ell, k)\,,
	\end{equation}
	where the vector $\bc(k)$ determines the linear combination
	(e.g., $\bc(k) = [1\,0\,\cdots\, 0]^{\T}$ selects the first channel).
	The \ac{RTF} vector, $\hr(k)$, is related to the transfer function vector,
	$\bh_\rd(k)$, by
	\begin{equation}
	\hr(k) = \frac{\bh_\rd(k) }{\bc^\H(k)\bh_\rd(k)} \,.
	\end{equation}
We refer to $\sr$ as the \emph{desired signal}.    

The signal processing system receives $\bx[n]$ [or equivalently $\bx(\ell, k)$] as an input and returns $\sest\;\![n]$, an estimate of the desired speech signal
$\sr[n]$,
as the output.  The estimate should effectively suppress noise and interference while maintaining low distortion and high intelligibility.  The algorithm should have low latency to facilitate real time operation, and should be able to adapt rapidly to changes in the scenario.
	
\section{Proposed algorithm}\label{sec:algorithm}
The various stages of the proposed algorithm are presented in this section.  The signal processing employs beamforming to suppress undesired components.  A speech detection method is used to determine which time-frequency bins are dominated by the desired speech and thus facilitate accurate estimation of the statistics used by the beamformer.  The beamformer's single-channel output undergoes post-processing to further reduce noise.  Initial calibration procedures
for the estimation of \acp{RTF} and sensor noise are  
also described.

\subsection{Beamforming framework}
A beamformer forms a linear combination of the input channels with the objective of enhancing the signal.  This operation can be described as
\begin{equation} \label{eq:applyweights}
y(\ell, k) = \bw^\H(\ell, k) \bx(\ell, k) \,,
\end{equation}
where $y(\ell, k)$ is the beamformer output, and $\bw(\ell, k)$ is the weight vector.  This can be in turn presented as a combination of filtered desired and undesired components
\begin{equation}
y(\ell, k) = \bw^\H(\ell, k) \hr(k) \sr(\ell, k) + \bw^\H(\ell, k) \bv(\ell, k)\,.
\end{equation}
The power of the undesired component at the beamformer's output is
\begin{equation}\label{eq:noise}
E\{ |\bw^\H(\ell, k) \bv(\ell, k)|^2 \} =  \bw^\H(\ell, k) \bPhi_{\bv \bv}(\ell ,k) \bw(\ell, k) \,,  
\end{equation}
where $\bPhi_{\bv \bv}(\ell ,k)= E\{ \bv(\ell, k) \bv^\H(\ell, k)  \}$.
The level of desired-signal distortion can be expressed as
\begin{equation}\label{eq:distortion}
|\bw^\H(\ell, k)\hr(\ell, k) - 1 |^2\,.
\end{equation}


There is a certain degree of 
discrepancy
between the dual objectives of reducing noise \eqref{eq:noise} and reducing distortion \eqref{eq:distortion}. The \ac{MVDR} beamformer minimizes noise under the constraint that \emph{no distortion} 
is allowed.  Formally,
\begin{align}\label{eq:optmvdr}
&\bw_{\mathrm{MVDR}}(\ell, k) = \argmin_{\bw(\ell, k)} \{ \bw^\H(\ell, k) \bPhi_{\bv \bv}(\ell ,k) \bw(\ell, k) \} \notag \\
&\textrm{   s.t.}     \quad\quad    \bw^\H(\ell, k)^\H \hr(k) =0  \,.
\end{align}
The solution to \eqref{eq:optmvdr} is
\begin{equation}\label{eq:mvdr}
\bw_{\mathrm{MVDR}}(\ell, k) =
\frac{\bPhi_{\bv\bv}^{-1}(\ell, k)\hr(k)}{\hr^\H(k) \bPhi_{\bv\bv}^{-1}(\ell, k)\hr(k)} \,.
\end{equation}

In contrast to \ac{MVDR} beamforming's constrained minimization of noise,
the \ac{MWF} preforms unconstrained minimization of the \ac{MSE} (i.e., distortion \emph{is} allowed). This leads to improved noise suppression but introduces distortion.
Formally, the \ac{MWF} is defined as
\begin{equation}
\bw_{\mathrm{MWF}}(\ell, k) =\argmin_{\bw(\ell, k)}
E\{|\bw^\H(\ell, k) \bx(\ell, k) - \sr(\ell, k)|^2 \}
\,.
\end{equation}
It has been shown \cite{Spriet04} that the \ac{MWF} is equivalent to 
performing
post-processing
to
the output of an \ac{MVDR} beamformer with a \acf{SWF}:
\begin{equation}
\bw_{\mathrm{MWF}}(\ell, k) = \bw_{\mathrm{MVDR}}(\ell, k) \cdot \WF(\ell, k)\,.
\end{equation}
An \ac{SWF}, $\WF(\ell, k)$, is determined from the \ac{SNR} at its input
(in this case the output of a beamformer).
The relationship is given by
\begin{equation}\label{eq:swf}
\WF(\ell, k) =\frac{1}{1 + \mathrm{SNR}^{-1}(\ell,k)}\,.
\end{equation}
We adopt this two-stage perspective and split the processing into an \ac{MVDR} stage followed by Weiner based post-processing.


For the \ac{MVDR} stage, knowledge of the \ac{RTF} $\hr(k)$ and of the noise covariance $\bPhi_{\bv\bv}(\ell, k)$ are required to compute the beamformer weights of \eqref{eq:mvdr}. The \ac{RTF} $\hr(k)$ can be assumed to remain constant since the positions of the sensors with respect to the mouth of the desired source are fixed.  Therefore, $\hr(k)$ can be estimated once during a calibration procedure and used during all subsequent operation. A framework for estimating the \ac{RTF} is outlined in Sec.~\ref{Sec:RTF}.

The noise covariance $\bPhi_{\bv\bv}(\ell, k)$ does not remain constant as it is influenced by changes in the user's position as well as changes of the characteristics of the undesired sources.  Therefore, $\bPhi_{\bv\bv}(\ell, k)$ must be estimated on a continual basis. The estimation of $\bPhi_{\bv\bv}(\ell, k)$ is described in Sec.~\ref{sec:noiseEst}.

The post-processing stage described in Sec.~\ref{sec:postprocessing} incorporates a scheme for estimating the \ac{SNR} of \eqref{eq:swf}.  Measures to limit the distortion associated  with Wiener filtering are also employed.

\subsection{Noise covariance estimation}\label{sec:noiseEst}
Since the noise covariance matrix $\bPhi_{\bv\bv}(\ell, k)$ may be subject to rapid changes, it must be continually estimated from the signal $\bx(\ell, k)$.  This may be accomplished by performing, for each frequency band, a weighted time-average which ascribes greater significance to more recent time samples.  As a further requirement, we wish to exclude bins which contain the desired speech component from the average, since their inclusion introduces bias to the estimate and is detrimental to the beamformer's performance. 
We estimate the noise variance as
\begin{equation} \label{eq:updatePhi}
\begin{aligned}
\bPhihat_{\bv\bv}(\ell, k) = \alpha(\ell, k)&\bPhihat_{\bv\bv}(\ell-1, k)\\ &+ \big(1 - \alpha(\ell, k)\big) \bx(\ell,k)\bx^\H(\ell,k) \,,
\end{aligned}
\end{equation}
where $\alpha$ is the relative weight ascribed to the previous estimate and $1 - \alpha$ is the relative weight of the current time instant.  If desired speech is detected during a given bin, $\alpha$ is set to 1, effectively ignoring that bin.  Otherwise, $\alpha$ is set to $\alpha_0 \in (0, 1)$.  Formally,
\begin{equation} \label{eq:setalpha}
\alpha(\ell, k) = 
\begin{cases}
    1,          & \text{if desired speech is detected}\\
    \alpha_0,   & \text{otherwise.}
\end{cases}
\end{equation}
The parameter $\alpha_0$ is a smoothing 
parameter 
which corresponds to a time-constant $\tau$ specifying the effective duration of the estimator's memory.  They are related by
\begin{equation}
  \alpha_0 = e^{\frac{-\tau F_s}{R}} \Leftrightarrow \tau = \frac{-R}{F_s \ln(\alpha_0)} \,,
\end{equation}
where $F_s$ is the sample rate, $R$ is the hop size (i.e., number of time samples between successive time frames), and $\tau$ is measured in seconds.  

In certain scenarios, $\bPhihat_{\bv\bv}(\ell, k)$ is ill-conditioned and \eqref{eq:mvdr} produces exceptionally large weight vectors \cite{Cox87}.  To counter this phenomenon, we constrain the norm of $\bw$ to a maximal value\footnote{It should be noted that \eqref{eq:regularization} is fairly rudimentary.  Other regularization methods which are more advanced exist such as diagonal loading \cite{Gilbert1955, Cox87}, alternative loading schemes \cite{Levin13robust}, and eigenvalue thresholding \cite{Harmanci00}.  We decided 
to use \eqref{eq:regularization} due to its computational simplicity: no weighting coefficient must be determined, nor is eigenvalue decomposition called for.},
\begin{equation} \label{eq:regularization}
\bw_{\mathrm{reg}}(\ell, k) =
\begin{cases}
    \bw_{\mathrm{MVDR}}(\ell, k),    & \text{if } \|\bw_{\mathrm{MVDR}}(\ell, k)\|_2 \le \rho \\
    \rho \frac{\bw_{\mathrm{MVDR}(\ell, k)}}{\|\bw_{\mathrm{MVDR}(\ell, k)}\|} ,   & \text{otherwise.}
\end{cases}
\end{equation}
In \eqref{eq:regularization}, $\bw_{\mathrm{reg}}$ represents the regularized weight vector and $\rho$ is the norm constraint.
	
\subsection{Narrowband near-field speech detection}
To determine whether the desired speech is present in a specific time-frequency bin, we propose the test statistic
\begin{equation}\label{eq:testnow}
T(\ell, k) = \frac{|\bx^\H(\ell, k) \hr(k)|^2}
 {|\bx(\ell, k)|^2 |\hr( k)|^2} \,.
\end{equation}
Geometrically, $T$ corresponds to the square of the cosine of the angle between the two vectors $\bx$ and $\hr$.  The highest value which $T$ may obtain is 1; this occurs when $\bx$ is proportional to $\hr$ corresponding to complete affinity between the received data $\bx$ and the \ac{RTF} vector $\hr$.
Speech detection is determined by comparison with a threshold value $\eta\,$:
for $T(\ell, k) \ge \eta$, speech is detected; otherwise, speech is deemed absent. This criterion determines the value of $\alpha(\ell, k)$ in \eqref{eq:setalpha}.

\subsection{Post-processing}\label{sec:postprocessing}
Post-filtering achieves further noise reduction at the expense of increased distortion.  Weiner filtering \eqref{eq:swf} applies an \ac{SNR} dependent attenuation (low \acp{SNR} incur higher levels of attenuation).
However, the \ac{SNR} is not known and needs to be estimated.  We use a variation of Ephraim and Malah's ``decision-directed'' approach \cite{Ephraim84}\red{, i.e.,}
\begin{equation}
\begin{aligned}
\gamma(\ell, k) &= \frac{|\bw_\mathrm{reg}^\H(\ell, k) \bx(\ell, k)|^2}
{\bw_\mathrm{reg}^\H(\ell, k)\bPhihat_{\bv\bv}(\ell, k)\bw_\mathrm{reg}(\ell, k)}   \\
\widehat{\mathrm{SNR}}(\ell, k) &=
\beta |\widehat{\WF}(\ell-1, k)|^2 \, \gamma(\ell-1, k)\\
&\quad \quad+(1 - \beta) \max\{\gamma(\ell, k)-1,\, \mathrm{\mathrm{SNR_{min}}}  \} \\
\widehat{\WF}(\ell, k) &=
\max\left\{\frac{1}{1 + \widehat{\mathrm{SNR}}^{-1}(\ell,k)}
    , \, \WF_{\mathrm{min}} \right\} \,.
\end{aligned}\
\end{equation}
The parameters $\mathrm{SNR_{min}}$ and $\WF_{\mathrm{min}}$ set thresholds that,
respectively, prevent the value of the estimated \ac{SNR} of the current sample and the amplitude of the Wiener filter from being overly low.  This limits the distortion levels and reduces the prevalence artifacts (such as musical tones).  However, this comes at the expense of less powerful noise reduction.  


Application of the post-processing stage to the output of the \ac{MVDR} stage yields the estimated speech signal:
	\begin{equation}\label{eq:applyfilt}
	\widehat{s}(\ell, k) = \widehat{\WF}(\ell, k)\, \bw^\H_\mathrm{reg}(\ell, k) \bx(\ell, k)\,.
	\end{equation}
The signal $\sest(\ell, k)$ which is in the time-frequency domain is then converted to the time domain to produce the system's final output, $\sest\;\![n]$.

\bed{It should be noted that the above approach is fairly rudimentary.
	Other more elaborate post-processing techniques have been developed and tested
	(e.g., \cite{McCowan03, Lefkimmiatis06, Gannot04}).
	Our simpler choice of post-processing algorithm serves for
	the purpose of demonstrating a complete system of beamforming and post-processing
	applied to a smartglasses system.  This concise algorithm could, in principal,
	be replaced with more sophisticated one.
	Furthermore, such algorithms could integrate information about speech activity contained in the test statistic $T(\ell, k)$.}
\begin{figure*}
	\centering
	\includegraphics[width=0.75\linewidth]{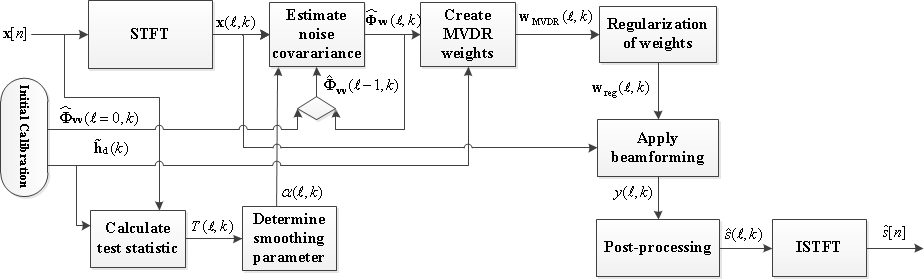}
	\caption{Block diagram schematic of the proposed algorithm.}
	\label{fig:BD}
\end{figure*}

\subsection{Calibration procedures} \label{sec:calibrate}
We describe calibration procedures which are done prior to running the algorithm.
\subsubsection{Sensor noise estimation}
Offline, the covariance matrix of sensor noise is estimated.  
This is done by recording a segment in which which only sensor noise is present.
Let the \ac{STFT} of this signal be denoted $\ba(\ell, k)$, and the 
number of time frames be denoted by ${L_{\mathrm{a}}}$.
Since sensor noise is assumed stationary, the time-averaged covariance matrix, $\overline{\bPhi}_{\ba\ba}(k) = 
\frac{1}{{L_{\mathrm{a}}}}\sum_{\ell = 1}^{L_{\mathrm{a}}}\ba(\ell, k)\ba^\H(\ell,k) $, serves as an estimate of the
covariance of sensor noise.

This is used in \eqref{eq:updatePhi} as the initial condition value $\bPhihat_{\bv\bv}(\ell-1, k)$.
Specifically, we set $\bPhihat_{\bv\bv}(\ell=0, k) = \overline{\bPhi}_{\ba\ba}(k)$.
It should be noted that setting the initial value as zeros would be problematic since this leads to a series of singular matrices.

\subsubsection{RTF estimation} \label{Sec:RTF}
System identification generally requires knowledge of 
a reference signal
and the system's output signals.
Let $\bb(\ell, k)$ represent the \ac{STFT} of speech signals produced by a user wearing the
glasses in a noise-free environment.  For \ac{RTF} estimation, the reference signal
is $\bc^\H(\ell, k) \bb(\ell, k)$ and the system's outputs are $\bb(\ell, k)$.
An estimate of the \ac{RTF} vector is given by
\begin{equation}\label{eq:estRTF}
\widetilde{\bh}_\mathrm{est}(k) = 
\frac{
\sum_{\ell = 1}^{L_{\mathrm{b}}} \bb(\ell, k) \bb^\H(\ell, k) \bc(\ell, k)}
{\sum_{\ell = 1}^{L_{\mathrm{b}}} |\bc^\H(\ell, k) \bb(\ell, k)|^2 } \,,
\end{equation}
where ${L_{\mathrm{b}}}$ denotes the number of time frames, and division is to be understood in an
element-wise fashion.

Since, the desired \ac{RTF} $\hr(k)$ is comprised only of the direct component, the input and output signals would ideally need to be acquired in an anechoic environment.  The availability of such a measuring environment is often not feasible, especially if these measurements are to be preformed by the end consumer.
This being the case, reliance on \eqref{eq:estRTF} can be problematic, since reverberation
is also incorporated into $\widetilde{\bh}_\mathrm{est}(k)$.
(We note that this information about reverberation in the training stage is not useful for the algorithm, since the during actual usage reverberation changes.)

To overcome the problem, we suggest a method based of \eqref{eq:estRTF}
in which the estimation of $\hr(k)$ may be conducted in a reverberant environment.  We propose that \ac{RTF} be estimated from measurements in which the speaker's position shifts during the course of the measurement procedure.  
Alternatively, we may apply \eqref{eq:estRTF} to 
segments recorded at different positions, and average the resulting  estimates of $\hr(k)$.
The rationale for this approach is that the reverberant components of the \ac{RTF} change with the varying  positions and therefore tend to cancel out.  The direct component, on the other hand is not influenced by the desired speaker's position.

\subsection{Summary of algorithm}
A block diagram depicting how the different components of the algorithm interrelate is presented in Fig.~\ref{fig:BD}.  The algorithm receives the multichannel input signal $\bx[n]$ and produces a single-channel output $\sest\;\![n]$.  Below, we describe the operations of the various blocks and refer the reader to the relevant formulas.


\begin{itemize}
	\item The \textbf{STFT} and \textbf{ISTFT} blocks perform the \acl{STFT} and inverse \acl{STFT}, respectively.  The main body of the algorithm operates in the time-frequency domain, and these conversion steps are required at the beginning and end of the process.
	\item The \textbf{Initial calibration} procedures \red{estimate} the \ac{RTF} and the initial noise matrix \red{(as} described in Sec.~\ref{sec:calibrate}\red{)}.
	\item The \textbf{Calculate test statistic} block calculates $T(\ell,k)$ from \eqref{eq:testnow}.
	\item The \textbf{Determine smoothing parameter} block determines the value of $\alpha(\ell, k)$ according to \eqref{eq:setalpha}.  The criterion $T(\ell,k) \ge \eta$ signifies speech detection.
	\item The \textbf{Estimate noise covariance} block calculates \eqref{eq:updatePhi}.
	\item The \textbf{Create MVDR weights} block calculates \eqref{eq:mvdr}.
	\item The \textbf{Regularization of weights} block calculates \eqref{eq:regularization}.
	\item The \textbf{Apply beamforming} block calculates \eqref{eq:applyweights}, producing a regularized \ac{MVDR} beamformer output.
	\item The \textbf{post-processing} block corresponds to Sec.~\ref{sec:postprocessing}.	 	
\end{itemize}


\section{Performance evaluation} \label{sec:evaluation}  
This section describes experiments conducted to evaluate the proposed algorithm.  The ensuing quantitative results are presented and compared to other algorithms.  The reader is referred to the associated website \cite{biuweb} in order to listen to various audio signals. 

\subsection{Recording setup}
First, we describe experiments which were conducted to evaluate the performance of the proposed algorithm.  To perform the measurements, we used two Microflown USP probes \cite{deBree03}.  Each of these probes consists of a pressure sensor and three particle-velocity sensors.  The physical properties of particle-velocity correspond to a dipole directivity pattern.

\begin{figure}
\centering
\includegraphics[width=0.7\linewidth]{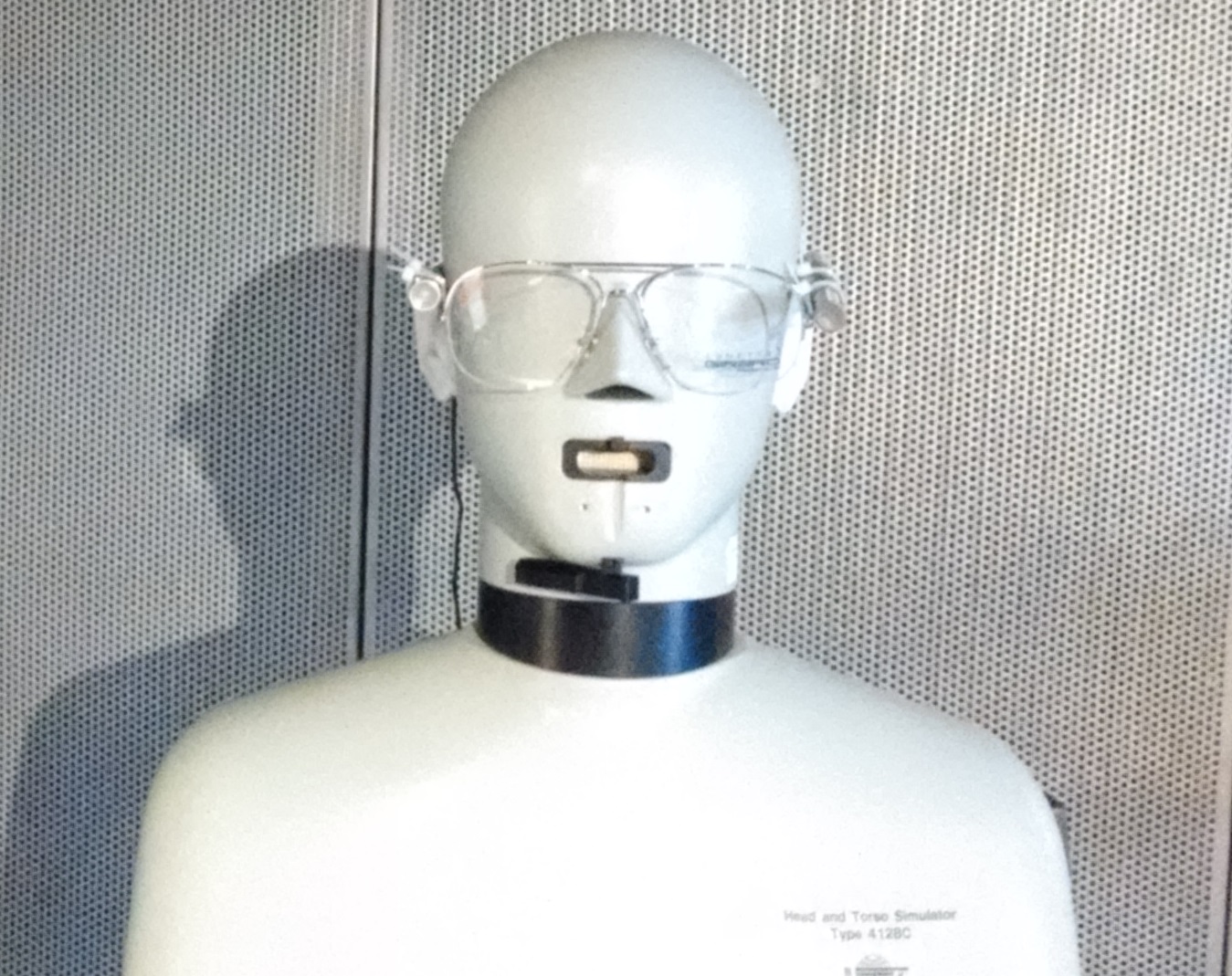}
\caption{Photograph of Cnut, our \acf{HATS},
	 eyeglasses array.
	 \bed{The distance from the center of Cnut's mouth to the \acp{AVS} is 
	 	approximately $10\frac{1}{2}$ cm.}
	 \label{fig:cnutPhoto}}
\end{figure}

The USP probes were fastened to the temples of eyeglasses\footnote{In 
this setup, the sensors are connected to external electronic equipment.
The recorded signals were processed afterwards on a PC.
The setup serves as a ``proof of concept" validation of the algorithm
preceding the development of an autonomous device.}
which were placed on the face of a \ac{HATS} (Br\"{u}el  Kj\ae{}r 4128C).   The \ac{HATS} is a dummy which is designed to mimic the acoustic properties of a human being.  For the sake of brevity, we refer to our \ac{HATS} by the nickname `Cnut'. 
\bed{The distance from the center of Cnut's mouth to the \acp{AVS} is 
	approximately $10\frac{1}{2}$ cm.}
 Fig.~\ref{fig:cnutPhoto} shows a photograph of Cnut wearing the glasses array.

Recordings of several acoustic sources were preformed in a controlled acoustic room.
These include the following five distinct voices:
\begin{enumerate}
\item A male voice, emitted from an internal loudspeaker located in Cnut's mouth.  This recording was repeated
four separate times, with changes made to Cnut's position or orientation between recordings.
Three of the recordings were used for \ac{RTF} estimation (as described in Sec.~\ref{Sec:RTF}).
The fourth recording was used to evaluate performance.  All other sources used for evaluation
(i.e., \#2--\#5) were recorded with Cnut in this fourth position and orientation.
\item A male voice emitted from an external static (i.e., non-moving) loudspeaker.
\item A female voice emitted from an  external static loudspeaker.
\item A male voice emitted from an  external static loudspeaker.
\item A male voice produced by one of the authors while walking in the acoustic room.
\end{enumerate}

\begin{figure}
\centering
\includegraphics[width=1.0\linewidth]{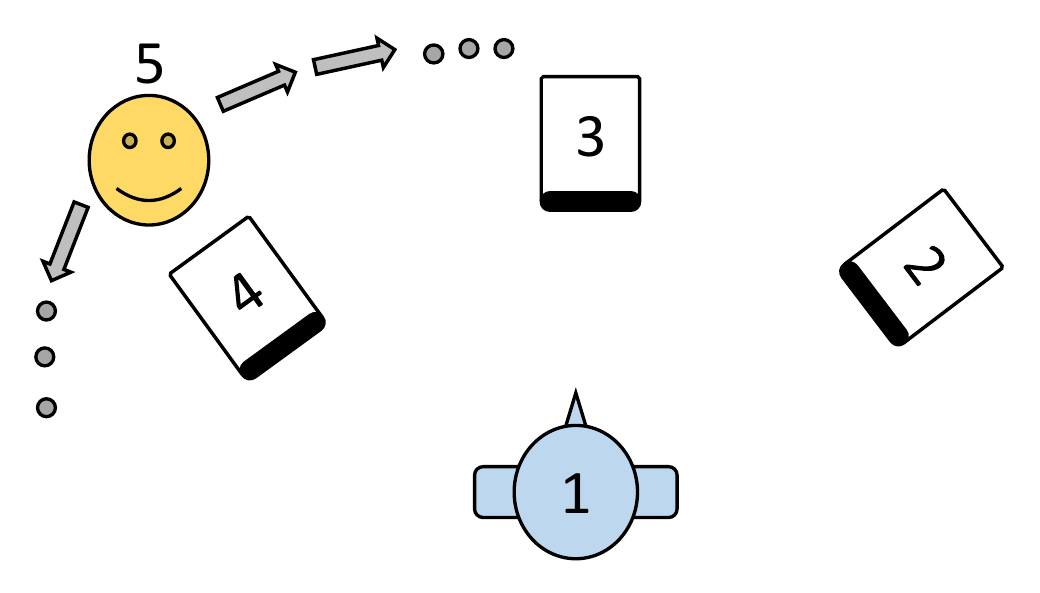}
\caption{Schematic of the acoustic sources. \#1: Cnut (the \ac{HATS});
	\#2--\#4: static loudspeakers; \#5: moving human speaker.
	\bed{The distance between the HATS and loudspeakers is approximately
	\red{1 meter}.}}
\label{fig:Schematic}
\end{figure}

These five separate voices were each recorded independently.  Sources \#2, \#3, and \#4 were located respectively to at the front-right, front, and front-left of Cnut and were positioned at a distance of approximately
\red{1 meter}
from it.  Source 5 walked along different positions of a semicircular path in the vicinity of sources \#1--\#3.
A rough schematic of the relative positions of the sources is given in Fig.~\ref{fig:Schematic}.

The recordings were conducted in the acoustic room at the speech and acoustics lab at
	Bar-Ilan University.  The room dimensions are approximately
	6 m $\times$ 6 m $\times$ 2.3 m.  During the recordings the reverberation level was
	medium-low.

The use of independent recordings for each source allows for the creation of more complex scenarios by forming linear combinations of the basic recordings.  These scenarios can be carefully controlled to examine the effects of different \acp{SNR}.  Furthermore, since both the desired speech and undesired speech components are known, we can inspect how the algorithm effects each of these.

The recordings were resampled from 48 kHz to 16 kHz.  Since the frequency-response of the USP sensors is not flat \cite{deBree03}, we applied a set of digital filters to equalize the channels.  These filters are also designed to counter nonumiformity of amplification across different channels and to remove low frequencies containing noise and high frequencies in the vicinity of 8 kHz (i.e., half the sampling rate of the resampled signals).

\begin{figure}
	\centering
	\includegraphics[width=0.95\linewidth]{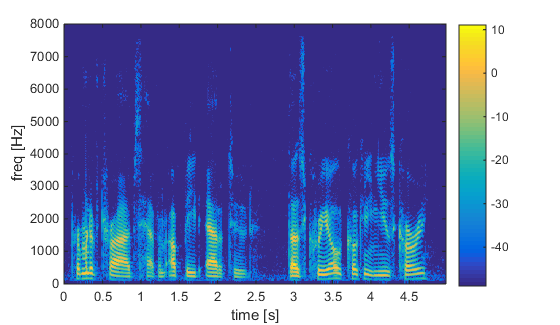}
	\caption{\red{A spectrogram depicting a segment of the speech signal emitted from
		Cnut's mouth as recorded by the monopole sensors.} 
		\label{fig:spect}}
\end{figure}

\bed{It should be noted that a sound wave with a frequency of 3.2 kHz has a wave length
	of approximately $10\frac{1}{2}$ cm, which corresponds to the distance between the center
	of Cnut's mouth and the \acp{AVS}.  For a typical speech signal, the bulk of the spectral
	energy in located beneath 3.4 kHz.
	Furthermore, the \red{power} spectrum of speech signals decays with increasing frequency \cite{Heute08}. 
	Both of these characteristics are qualitatively evident in 
	Fig.~\ref{fig:spect}, which portrays the spectrogram of a segment of the signal emitted from Cnut's
	mouth as recorded by the monopole sensors.  (This signal serves as the
	clean signal for evaluation purposes in Sec.~\ref{sec:perfsub}).}
\red{Accordingly, the vast majority of the desired speech signal's power
	corresponds to sub-wavelength propagation.}

\subsection{Processing details}
The calibration for sensor-noise was done with silent segments of the recordings and the \ac{RTF} estimation was done with recordings of speaker \#1 at different positions and orientations
and no other speakers mixed in.
The average of the two omnidirectional channels was used as the reference signal $\sr(\ell, k)$ for estimating the \ac{RTF} $\hr(k)$.
This corresponds to designating $\bc$ of \eqref{eq:cvec} as
$\frac{1}{2}[1\,0\,0\,0\,1\,0\,0\,0]$.

\red{The input signals upon which processing is preformed correspond to two 
	scenarios.}
In the first scenario, three static 
speakers (\#2--\#4) were combined with equal levels of 
mean power.  
Afterwards, they were added to the desired source (\#1) at different \ac{SNR} levels.  In the second scenario, source \#5 was combined with source \#1 at different \ac{SNR} levels.

Each of the 8 channels is converted to the time-frequency domain by the \ac{STFT}, and processed with the algorithm proposed in Sec.~\ref{sec:algorithm}.  
Presently, the post-processing stage is omitted; it is evaluated separately in Sec.~\ref{sec:evalpost}.  An inverse \ac{STFT} transform converts the output back into the time domain.  The values for the parameters used are specified in Table~\ref{table:params}.

\begin{table}
\caption{Parameter values used for testing proposed algorithm.} \label{table:params}
\rowcolors{1}{}{lightgray}
\begin{tabular}{l l}
 \textbf{Parameter:}            & \textbf{Value:} \\ \hline
  sampling frequency            & $f_\mathrm{s} = 16$ kHz                     \\ 
  analysis window               & 512 sample Hamming window   \\
  hop size                      & $R = 128$ samples  \\
  \shortstack[l]{FFT size\\ \phantom{()}}                      & \shortstack[l]{1024 samples\\$\;$ (due to zero-padding)} \\
  synthesis window              & 512 sample Hamming window \\
  \shortstack[l]{smoothing parameter\\ \phantom{()}}      & \shortstack[l]{$\alpha_0 = 0.98$ (corresponds to\\ $\;$ $\tau \approx 0.396$ seconds)} \\
  norm constraint               & $\rho = 15$\\
  \shortstack[l]{speech detection\\ $\;$threshold}    & $\eta = 0.9$ \\
\end{tabular}
\end{table}   

Several other algorithms are also examined as a basis for comparison.  These algorithms include
variations of \ac{MVDR} and \ac{MPDR} beamforming and are described below:
\begin{enumerate}
	\item \emph{Fixed-\ac{MVDR}} which uses a training segment in which only undesired components are present (i.e., prior to the onset of the desired speech) in order to 
	calculate the sample covariance-matrix $\overline{\bPhi}_{\bv\bv}(k)$
	which serves as an
	estimate of the noise  
	covariance matrix.  This matrix $\overline{\bPhi}_{\bv\bv}(k)$
	is estimated only once and then used for the duration of the processing. 
	\item \emph{Fixed-\ac{MPDR}} which uses the sample-covariance matrix
	calculated from the segment to be processed. 
	In \ac{MPDR} beamforming \cite{VanTrees02}, both desired and undesired signals
	contribute to the covariance matrix. The matrix, $\bPhi_{\bv\bv}$ is replaced by
	$\overline{\bPhi}_{\bx\bx}(k)$.
	In contrast to the fixed-\ac{MVDR} algorithm, no separate training segment is used.
	Instead, the covariance matrix is estimated once from the \emph{entire} segment to be processed,  
	and used for the entire duration of the processing.
	\item \emph{Adaptive-\ac{MPDR}} which 
	uses a time-dependent estimate of the covariance matrix of the received signals, $\bPhihat_{\bx\bx}(\ell, k)$. 
	This is done by running the proposed algorithm with the threshold parameter set at $\eta = 1$,
	effectively ensuring that $\alpha(\ell,k) = \alpha_0$ for all $\ell$ and $k$. 
	\item \emph{Oracle adaptation} \ac{MVDR}
	which uses a time-dependent estimate of the noise covariance matrix,
	$\bPhihat_{\bv\bv}(\ell, k)$, based on the pure noise-component  
	[i.e, $\bx(\ell, k)$ of  \eqref{eq:updatePhi} is replaced by the undesired component $\mathbf{v}(\ell, k)$, and $\eta$ is set to 1].  
	The pure noise component is unobservable in practice, hence the denomination `oracle';
	the algorithm is used for purposes of comparison.
	
	\item The \emph{unprocessed signal} (i.e., the average of the two omnidirectional sensors) is used for comparison with respect to the \ac{STOI} measure.  By definition, the noise reduction for the unprocessed signal is 0 dB and distortion is absent. 
\end{enumerate}

\bed{All algorithms use all eight channels of data.  Furthermore, we
	\red{apply} the proposed algorithm and the oracle adaptation
	\red{\ac{MVDR} to} 
	the data from a reduced array containing only the two
	monopole channels\footnote{\red{Estimations of the noise covariance matrix and the \ac{RTF} vector used for beamforming and estimation of $T(\ell, k)$ are then 
			of reduced size, being based on only two channels.}}.
	\red{The obtained results} provide an indication of the performance enhancement
	due to the additional six dipole channels present in the \acp{AVS}.}

The feasibility of using these algorithms in practical settings varies.  The fixed-\ac{MVDR} algorithm presumes that segments containing only noise are available.  The fixed-\ac{MPDR} algorithm requires knowledge of the entire segment to be processed and hence cannot be used in real-time.  Both the adaptive \ac{MPDR} and the proposed algorithm are capable of operation in real-time. The oracle adaptation is not realizable since pure undesired components are unobservable in practice.  Its function is purely for purposes \red{of comparison}.


\subsection{Performance}\label{sec:perfsub}
In this subsection, we conduct an analysis of the proposed algorithm's performance and compare it to the performance of other algorithms.  Three measures are examined: (i) noise reduction (i.e., the amount by which the undesired component is attenuated); (ii) distortion (i.e., the amount by which the desired component of the output differs from its true value); (iii) the \ac{STOI} measure \cite{Taal11}.

Each of these three measures entails comparing components of the processed signal to some baseline reference.  Since the signals at the eight different channels possess widely different characteristics, an arbitrary choice of a single channel to serve as the baseline, may produce misleading results.
For instance, if a signal originates from the side of the array, the user's head will shield some of the sensors.
Selecting a signal from the right temple as opposed to the left temple to serve as the baseline may produce very different results.  We elect to  use the average of 
powers
at the two omnidirectional sensors to define a signal's 
power.
In a similar fashion, the clean reference  signal is taken as the average of the desired components received at the two omnidirectional sensors (which should be similar due to the geometrical symmetry of the corresponding direct paths).
These definitions are stated formally below [\eqref{eq:def_nr}--\eqref{eq:def_clean}].
It should be stressed that this procedure relates to the \emph{evaluation} but has 
no impact on the \emph{processing} itself.

The procedure for testing an algorithm is as follows.  An algorithm normally receives the data $\bx[n]$, calculates the weights $\bw_{\mathrm{alg}}(\ell, k)$, and applies them to produce the output $\sest_{\mathrm{alg}}[n]$.  In our controlled experiments, the desired and noise components of $\bx[n]$ are known.  Hence, we can apply $\bw(\ell, k)$ to the desired and noise components producing $d_\mathrm{alg}[n]$ and $v_\mathrm{alg}[n]$, respectively.
Let us define $\bw_\mathrm{L}(\ell, k)$ and $\bw_\mathrm{R}(\ell, k)$ as weights which select only the left and right omnidirectional channel (i.e., the weight value for the selected channel is 1, and all other channel weights are 0 valued).
In a similar manner, $d_\mathrm{L}[n]$, $d_\mathrm{R}[n]$, $v_\mathrm{L}[n]$, and $v_\mathrm{R}[n]$ are produced.
The \emph{noise reduction} is defined as
\begin{equation}\label{eq:def_nr}
  \text{noise reduction} = \frac{\frac{1}{2}\sum_n (v_\mathrm{L}^2[n] + v_\mathrm{R}^2[n])}{\sum_n v_\mathrm{alg}^2[n]}\,.
\end{equation}
The distortion level is defined as
\begin{equation}\label{eq:def_dl}
  \text{distortion} = \frac{\sum_n \left( d_\mathrm{alg}[n] -  \frac{1}{2}(d_\mathrm{L}[n] + d_\mathrm{R}[n])\right)^2}
  {\sum_n \left(\frac{1}{2}(d_\mathrm{L}[n] + d_\mathrm{R}[n])\right)^2}\,.
\end{equation}
For calculating the \ac{STOI} level, we compare the algorithm's output $\sest_{\mathrm{alg}}[n]$
with 
\begin{equation}\label{eq:def_clean}
\mathrm{clean}[n] = \tfrac{1}{2} (d_\mathrm{L}[n] + d_\mathrm{R}[n])
\end{equation}
functioning as the clean reference signal.

\begin{figure*}
	\centering
	\begin{minipage}[b]{0.9\linewidth}
		\subfigure[3 static 
		interferers \label{fig:resultsNa}]{
			\includegraphics[width = 0.46\linewidth]{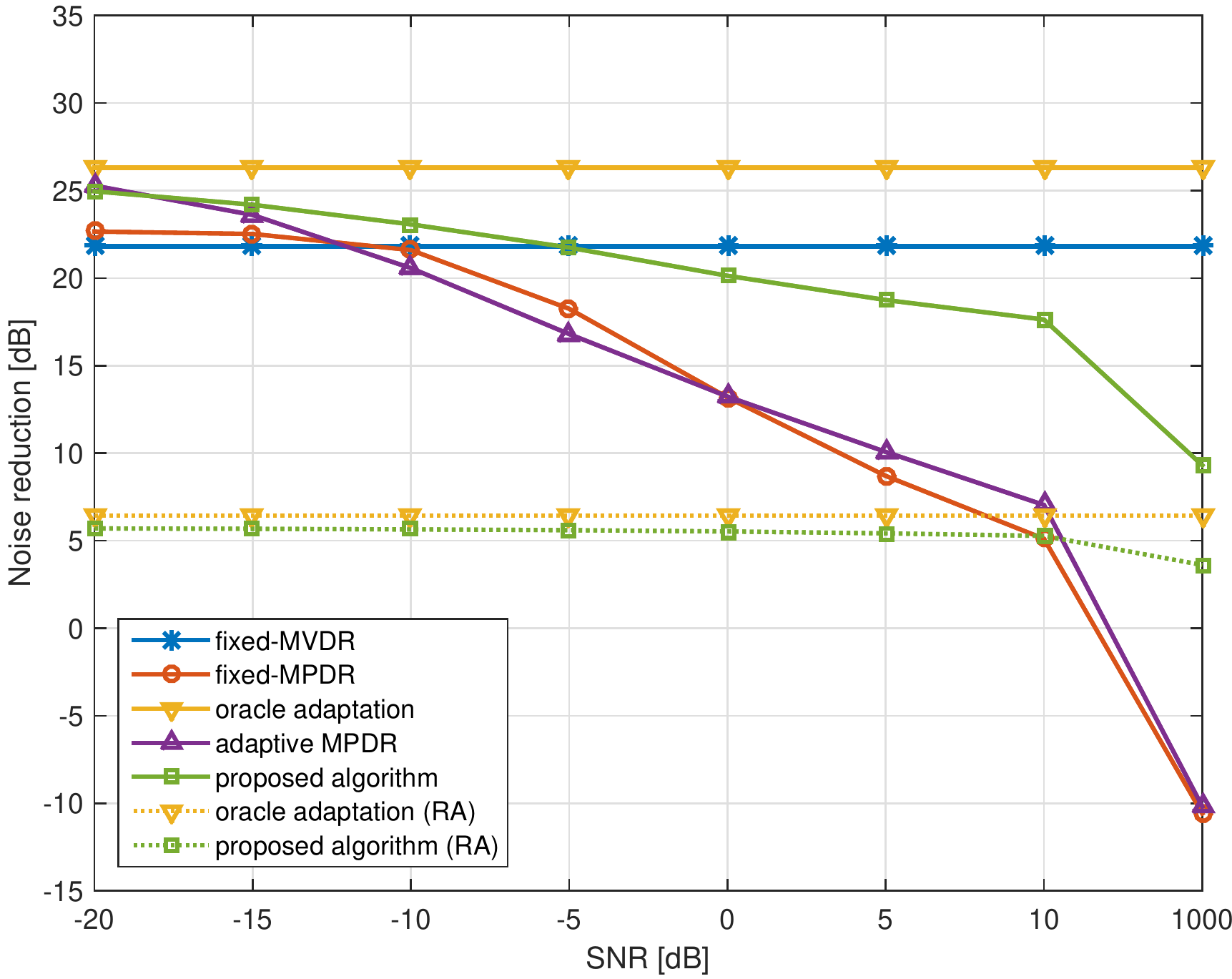}%
		}
		\subfigure[1 moving interferer \label{fig:resultsNb}]{
			\includegraphics[width = 0.46\linewidth]{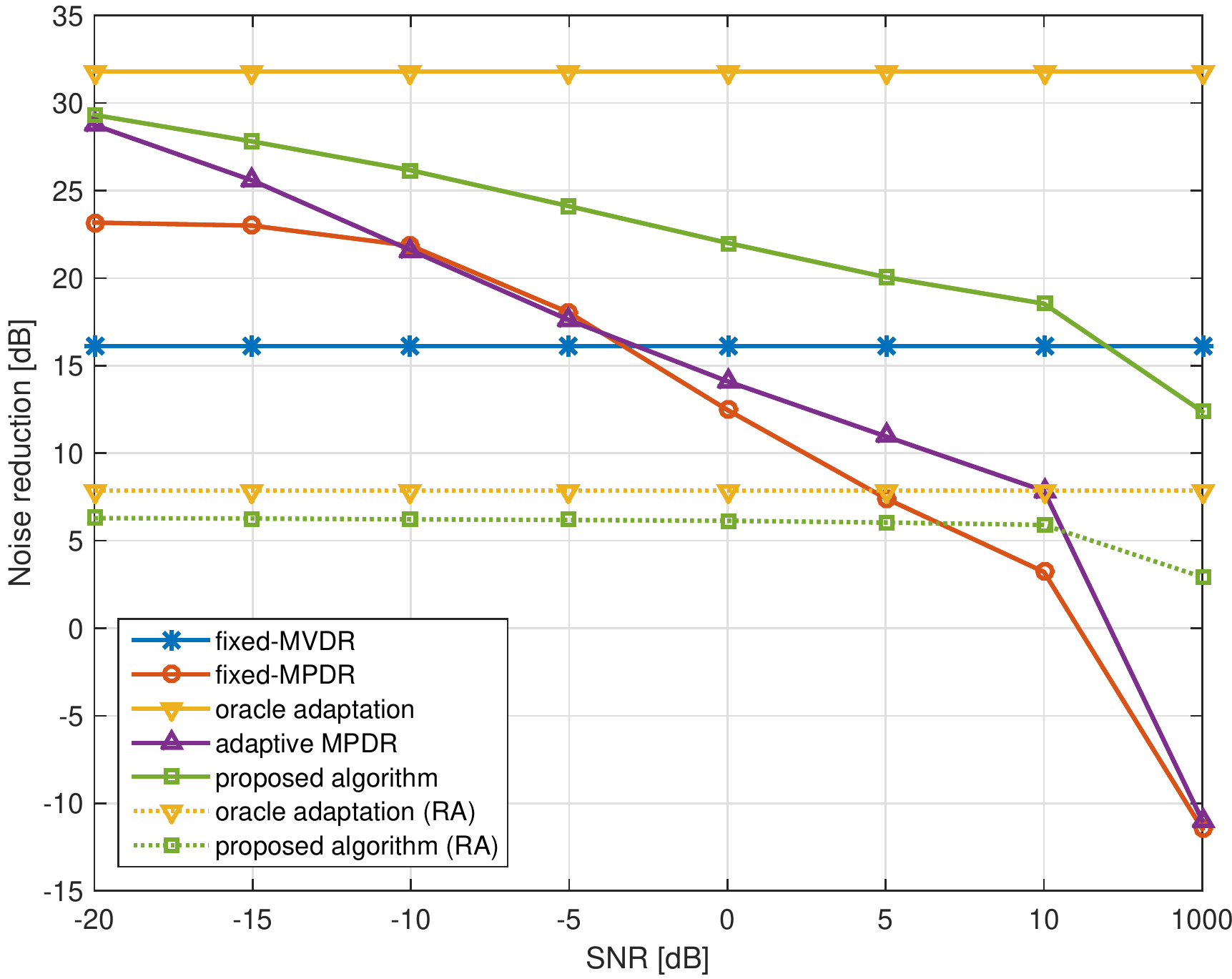} %
		}\caption{Noise reduction attained from processing with five different  algorithms
		for varying \ac{SNR} levels in two scenarios.}\label{fig:resultsN}%
	\end{minipage}\vspace{6mm}
	\begin{minipage}[b]{0.9\linewidth}
		\subfigure[3 static 
		interferers \label{fig:resultsDa}]{
			\includegraphics[width = 0.46\linewidth]{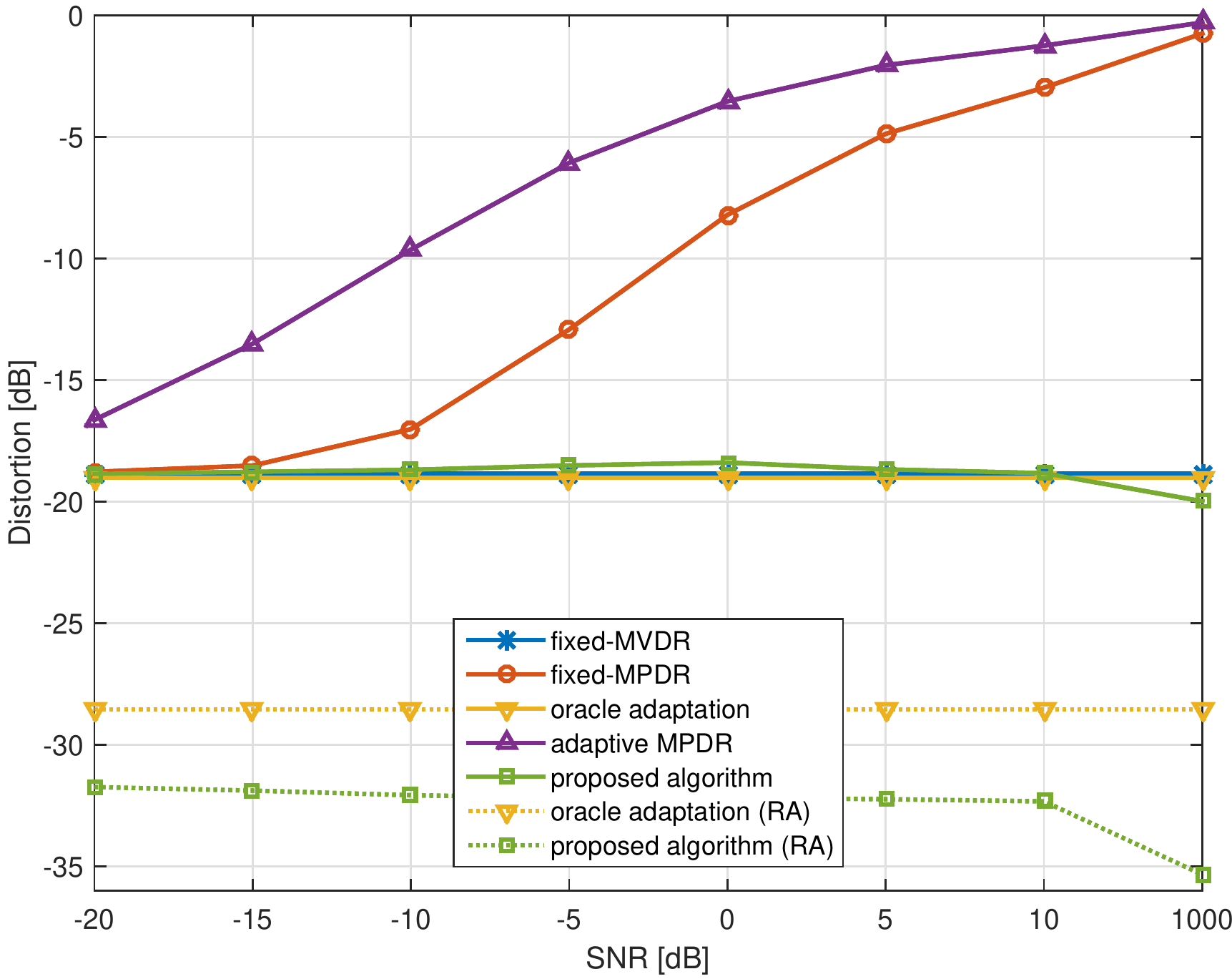}%
		}
		\subfigure[1 moving interferer]{
			\includegraphics[width = 0.46\linewidth]{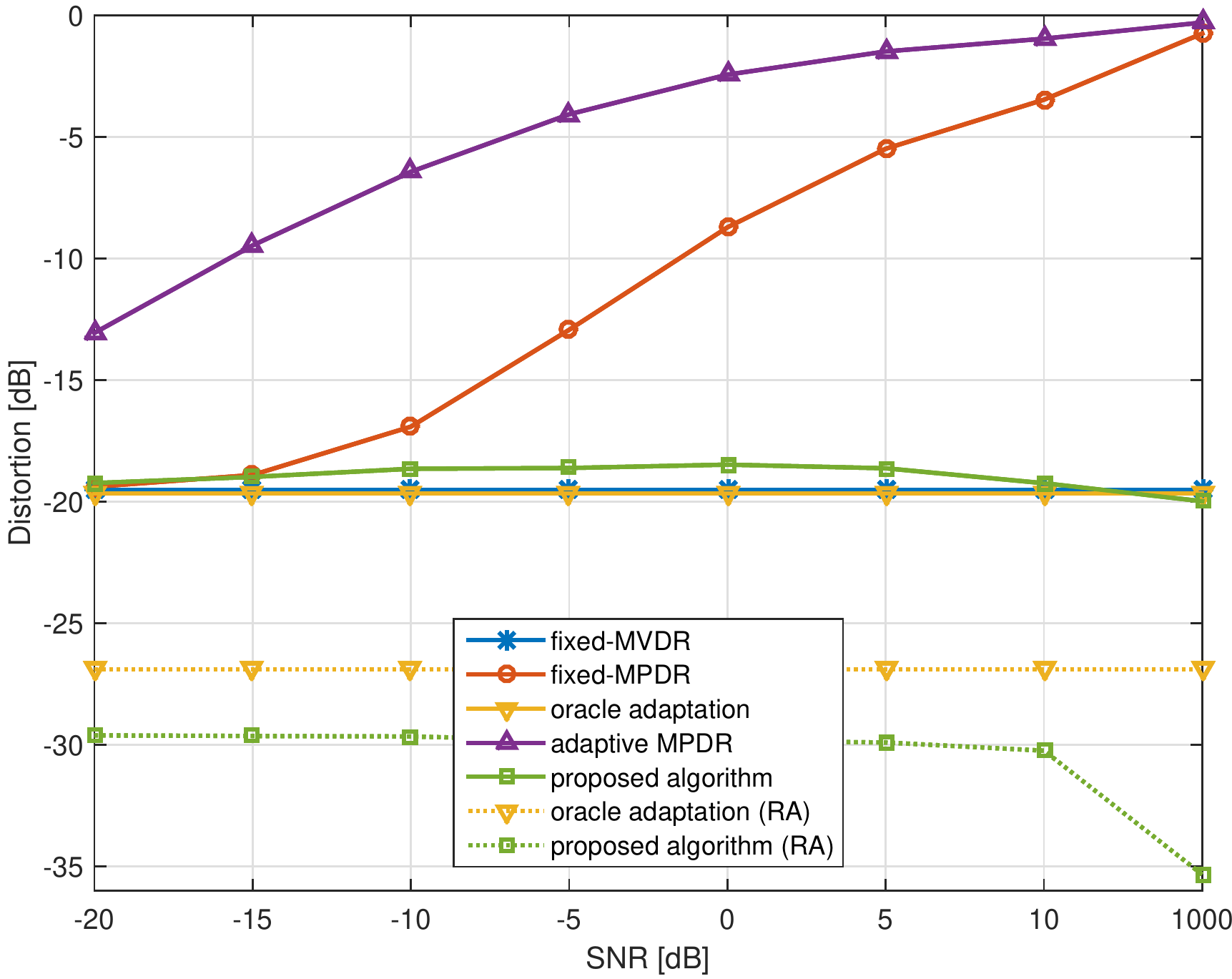}%
		}\caption{Distortion levels resulting from processing with five different algorithms
				for varying \ac{SNR} levels in two scenarios.}\label{fig:resultsD}
	\end{minipage}\vspace{6mm}
	\begin{minipage}[b]{0.9\linewidth}
		\subfigure[3 static 
		interferers \label{fig:resultsSa}]{
			\includegraphics[width = 0.46\linewidth]{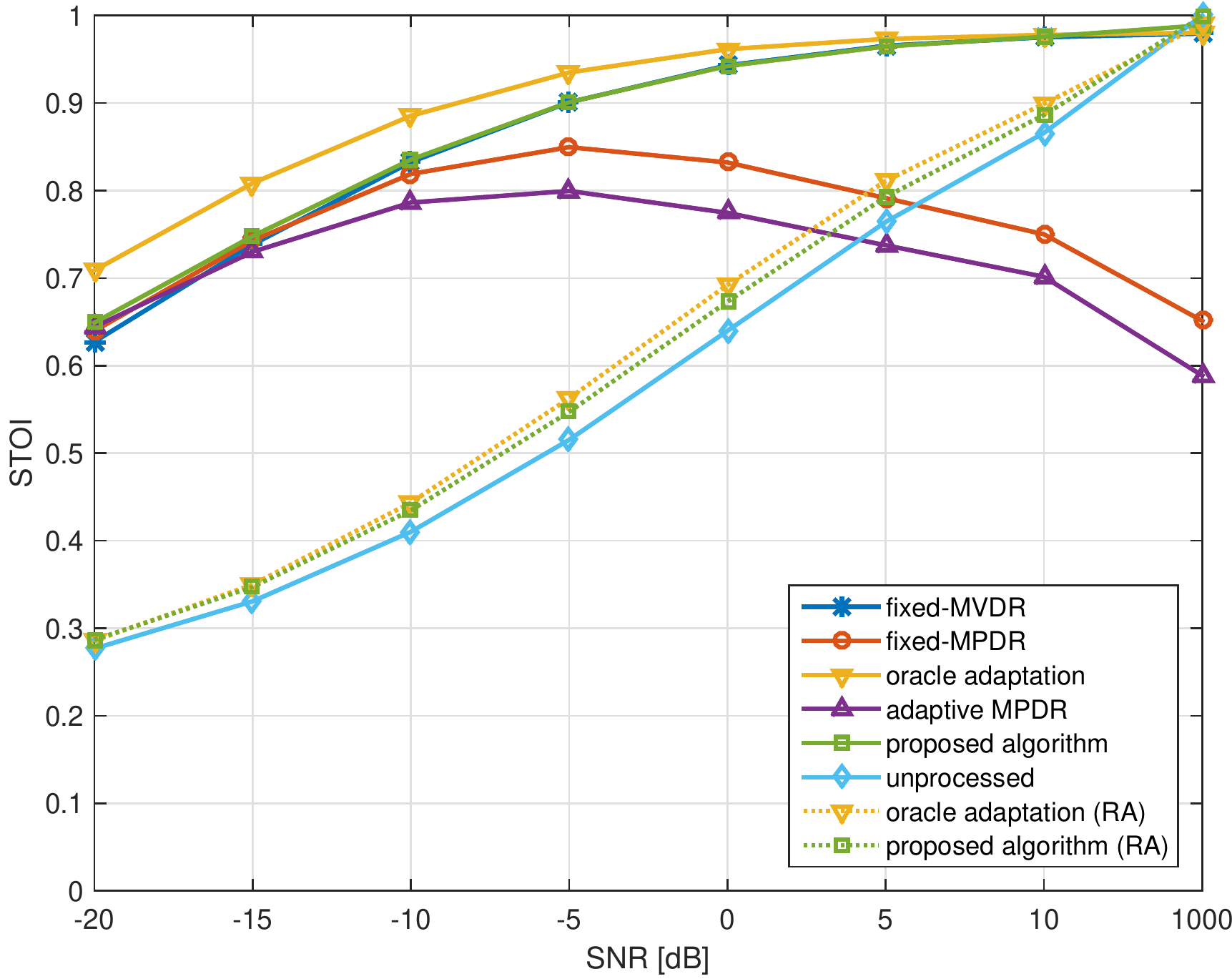}%
		}
		\subfigure[1 moving interferer]{
			\includegraphics[width = 0.46\linewidth]{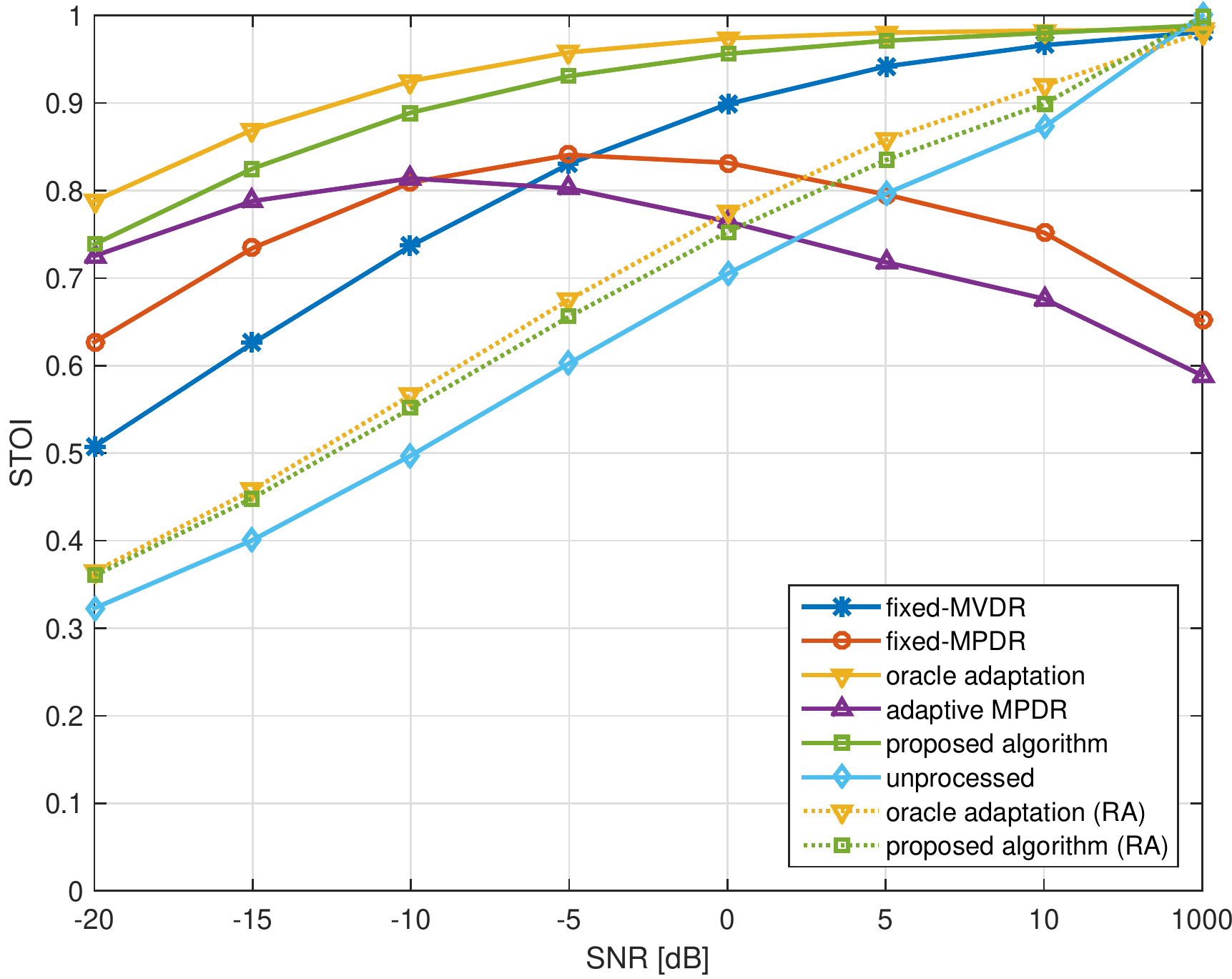}%
		}\caption{\ac{STOI} levels resulting from processing with five different  algorithms
				for varying \ac{SNR} levels in two scenarios.
				(Note that for the static scenario (a), the fixed-MVDR and the proposed algorithm are nearly identical.)}\label{fig:resultsS}
	\end{minipage}
\end{figure*}

\bed{We proceed to analyze the results of the algorithms
	\red{under test} 
	which utilize all eight channels 
	(these are marked by solid lines).  Afterwards, we return to the results which use only data from a \emph{reduced array} (marked in a dotted line) and compare.}
Fig.~\ref{fig:resultsN} portrays the noise reduction results, Fig.~\ref{fig:resultsD} portrays the distortion results, and Fig.~\ref{fig:resultsS} portrays the \ac{STOI} results.  The \acp{SNR} examined range from -20 dB to 10 dB with increments of 5 dB; furthermore an  \ac{SNR} of 1000 dB is also examined (appearing at the right edge of the horizontal axis).  This exceptionally high \ac{SNR} is useful for checking robustness in extreme cases.

\begin{figure*}
	\centering
	\begin{minipage}[b]{0.9\linewidth}
		\subfigure[3 \ed{static} interferers \label{fig:paramNa}]{
			\includegraphics[width = 0.46\linewidth]{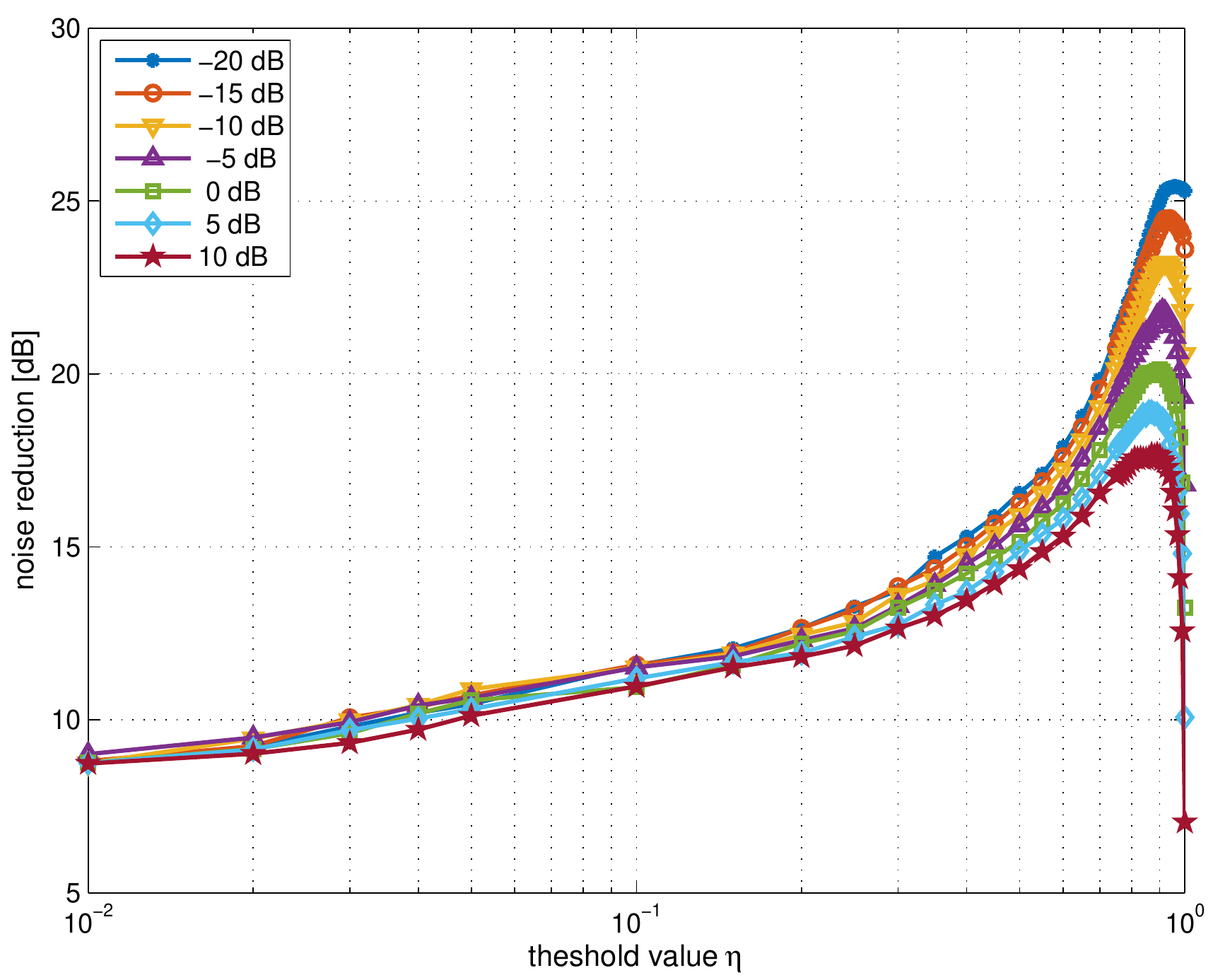}%
		}
		\subfigure[1 moving interferer \label{fig:paramNb}]{
			\includegraphics[width = 0.46\linewidth]{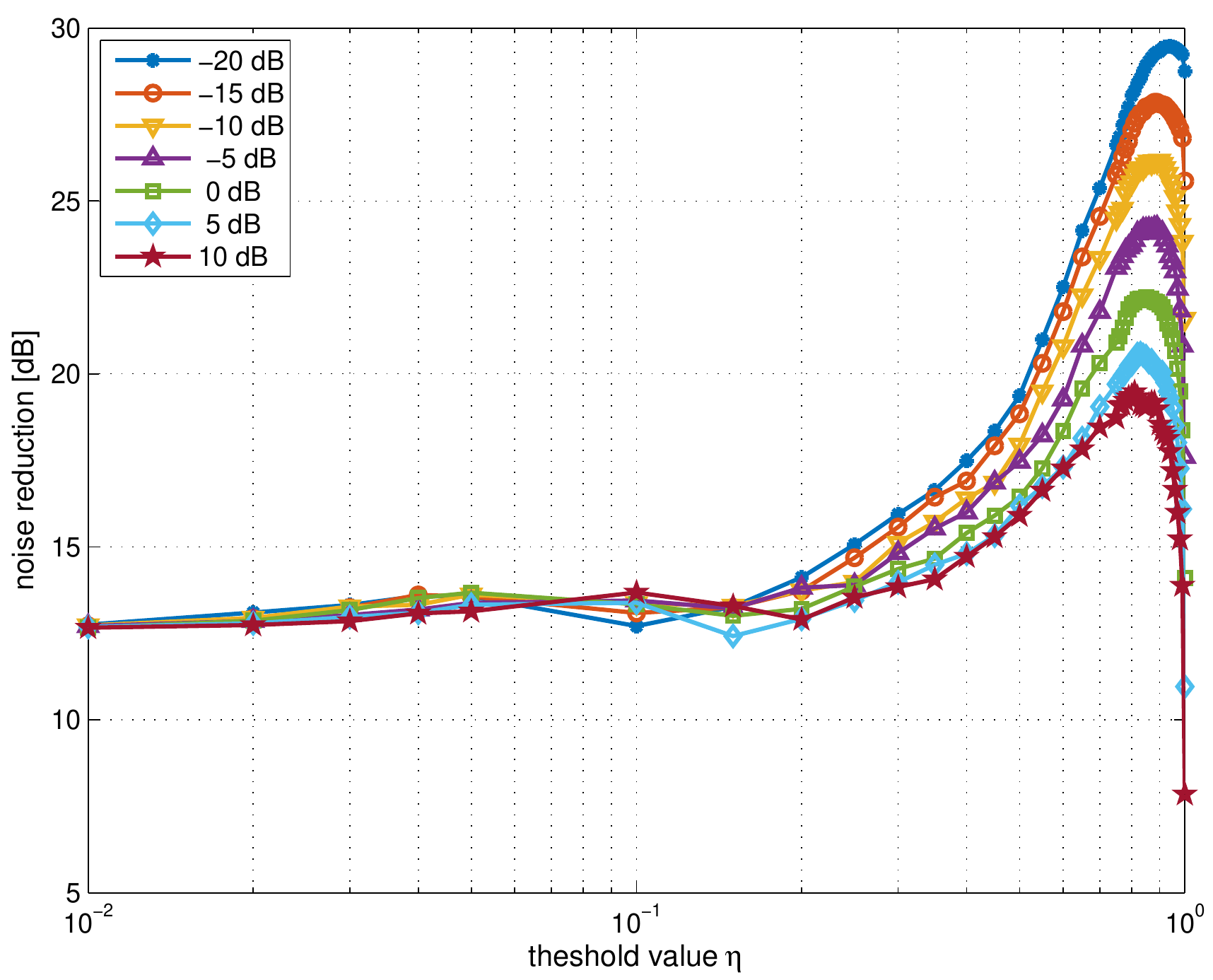} %
		}\vspace{-3mm}\caption{Noise reduction attained with the proposed algorithm using different values for the speech detection threshold ($\eta$) for a number of \ac{SNR} levels.}\label{fig:paramN}%
	\end{minipage}\vspace{6mm}
	\begin{minipage}[b]{0.9\linewidth}
		\subfigure[3 \ed{static} interferers]{
			\includegraphics[width = 0.46\linewidth]{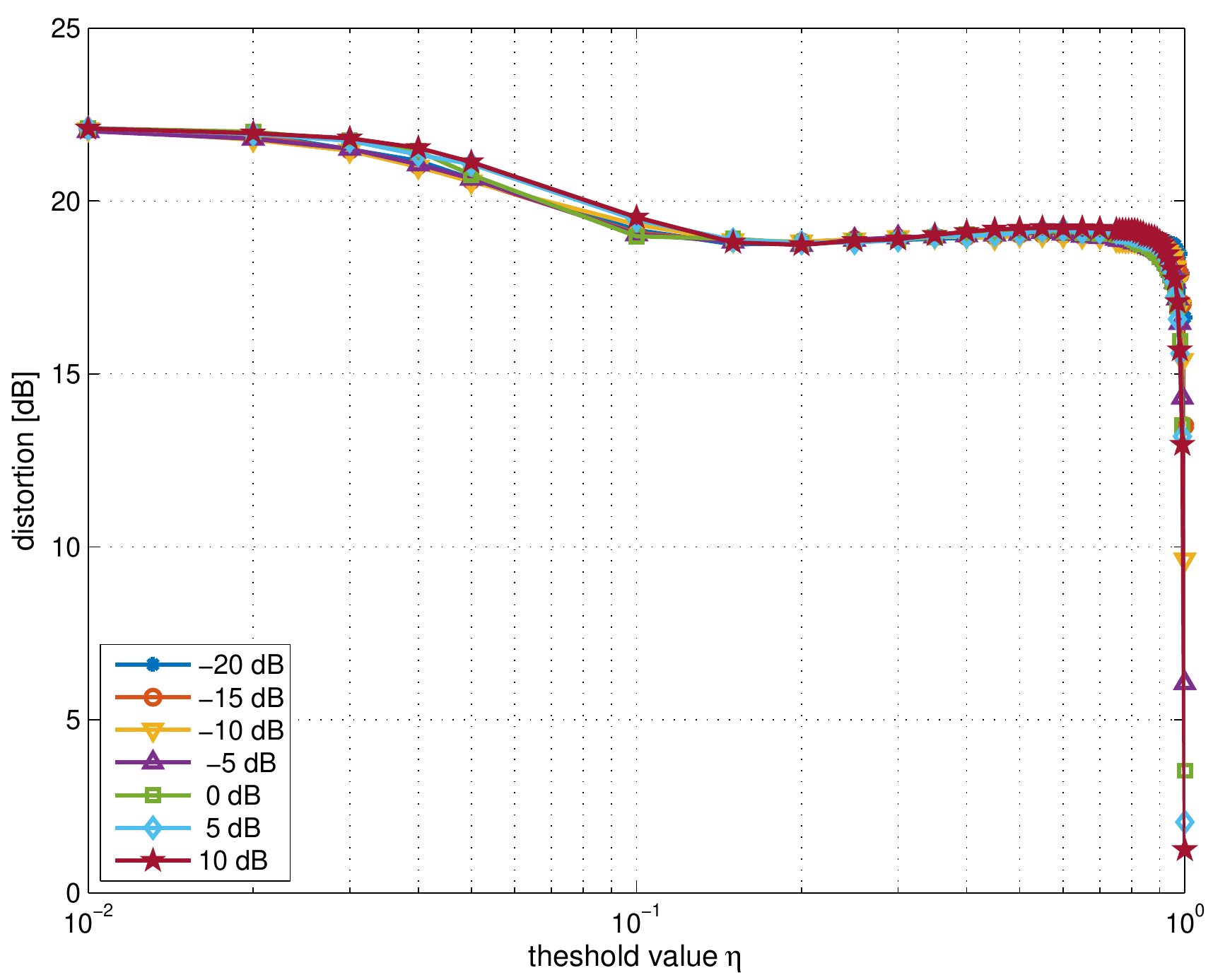}%
		}
		\subfigure[1 moving interferer]{
			\includegraphics[width = 0.46\linewidth]{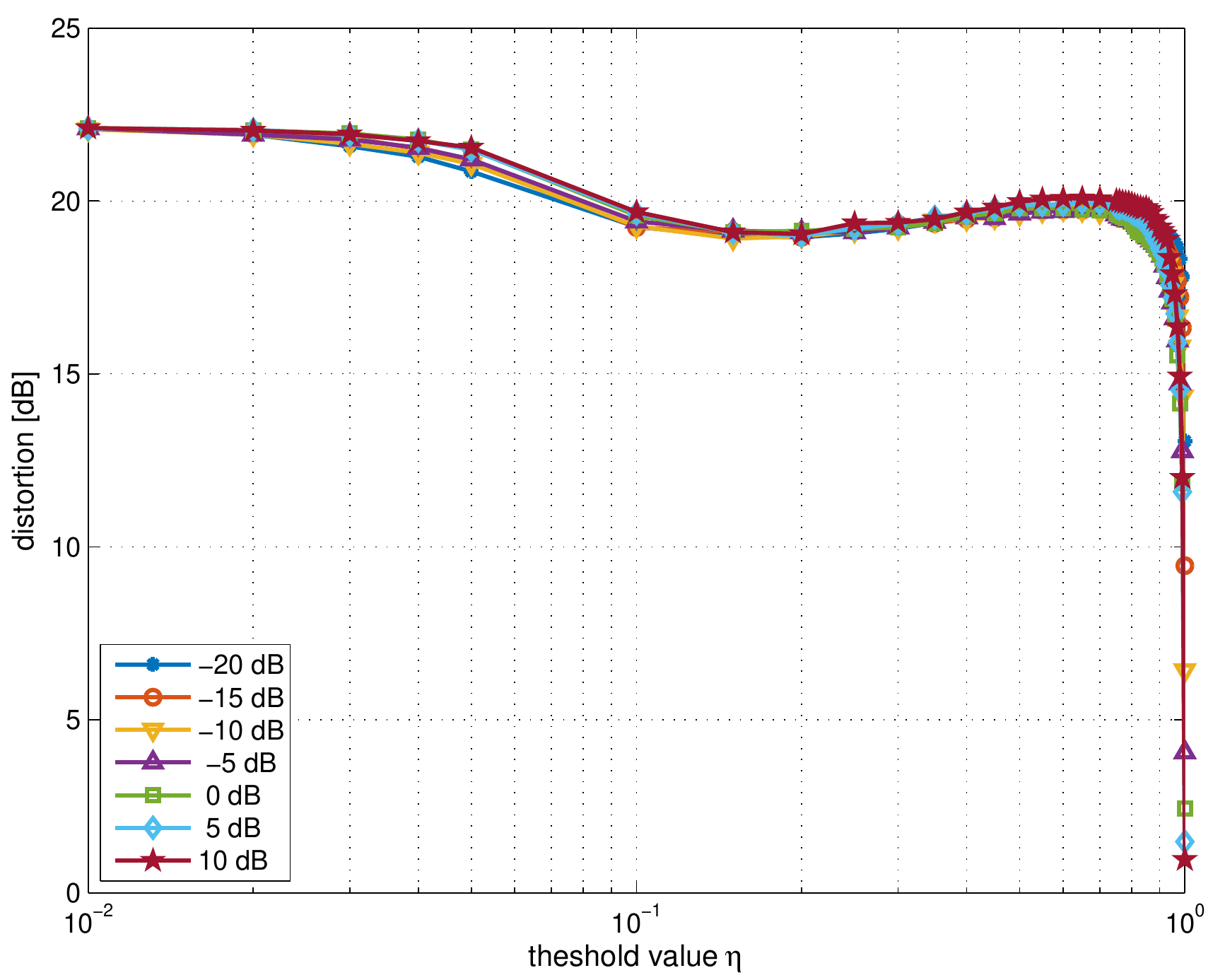}%
		}\vspace{-3mm}\caption{Distortion levels resulting from applying the proposed algorithm using different values for the speech detection threshold ($\eta$) for a number of \ac{SNR} levels.}\label{fig:paramD}
	\end{minipage}\vspace{6mm}
	\begin{minipage}[b]{0.9\linewidth}
		\subfigure[3 \ed{static} interferers]{
			\includegraphics[width = 0.46\linewidth]{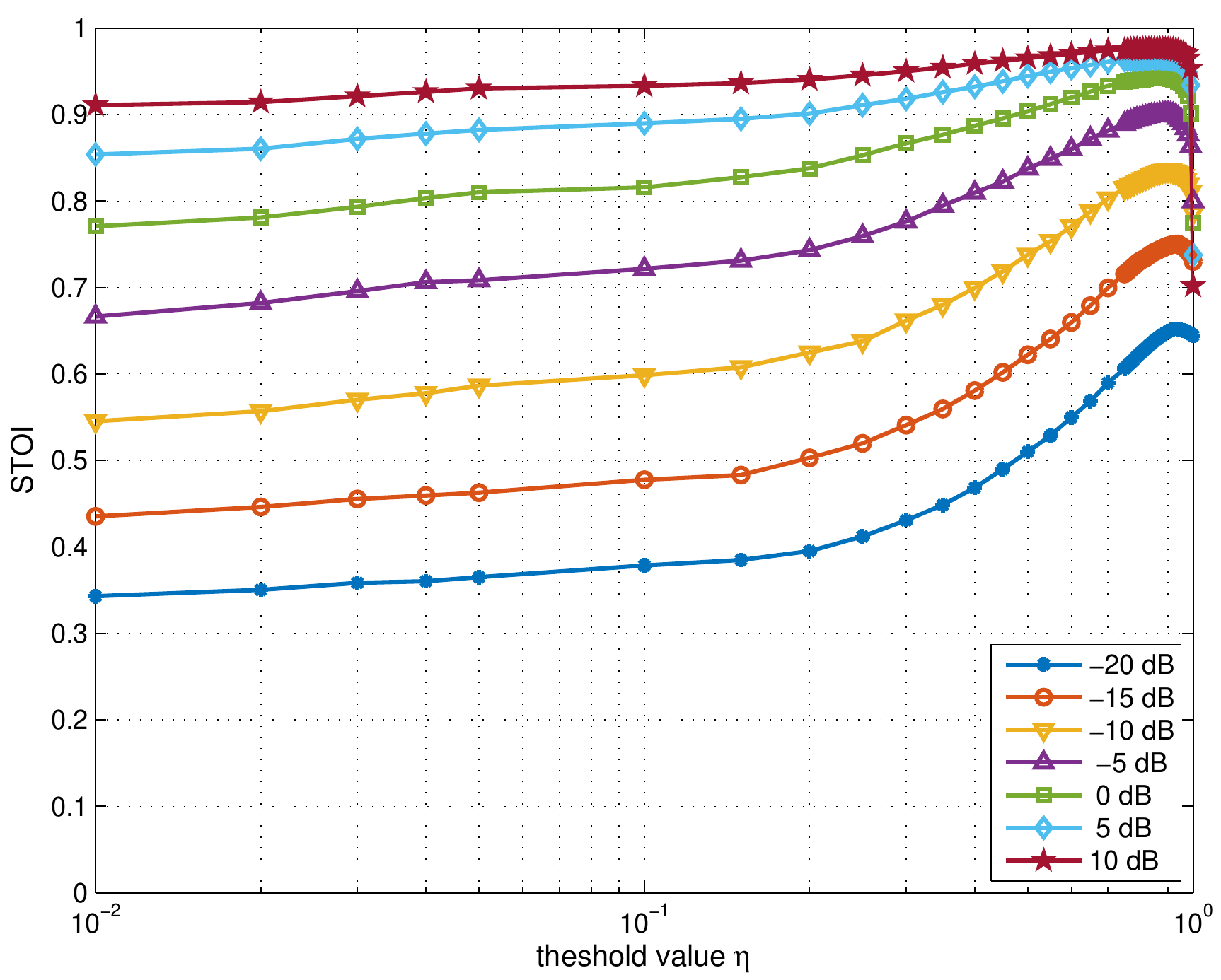}%
		}
		\subfigure[1 moving interferer]{
			\includegraphics[width = 0.46\linewidth]{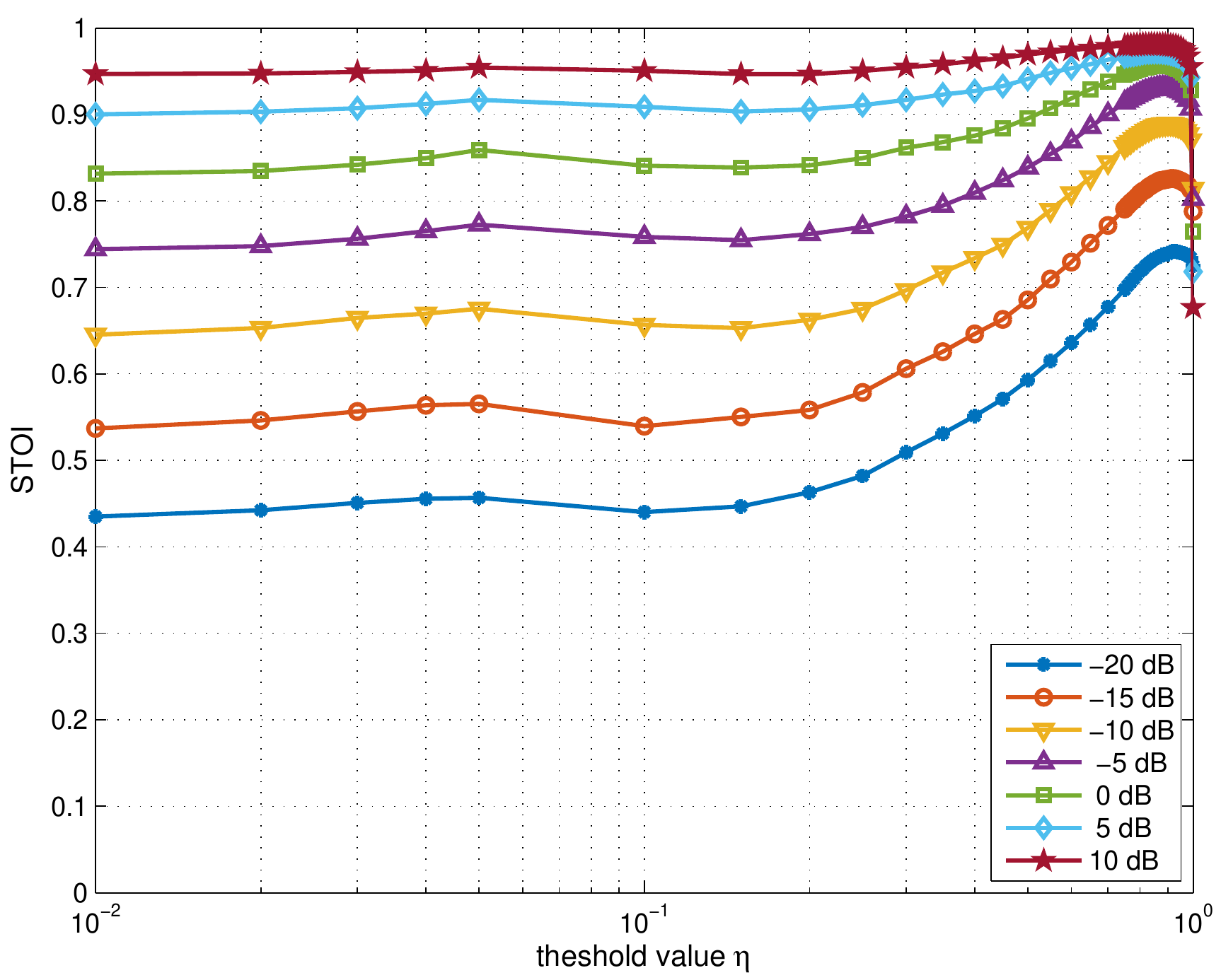}%
		}\vspace{-3mm}\caption{\ac{STOI} levels attained with the proposed algorithm using different values for the speech detection threshold ($\eta$) for a number of \ac{SNR} levels.}\label{fig:paramS}
	\end{minipage}
\end{figure*}

The results shown in the figures indicate that although both \ac{MPDR} based algorithms perform reasonably well for low \acp{SNR}, there is a rapid degradation in performance as \ac{SNR} increases.  This can be explained by the contamination of the estimated covariance-matrix by desired speech, which is inherent in these methods.  For very low \acp{SNR} the contamination is negligible, but at higher \acp{SNR} the contamination becomes significant.  Due to this issue, the \ac{MPDR} based algorithms cannot be regarded as viable.

With respect to distortion, the other algorithms (i.e., fixed-\ac{MVDR} and proposed) score fairly well with levels between $-20$ dB and $-18$ dB.  However, they differ with regards to noise reduction.  For the static scenario, the fixed-\ac{MVDR} attains a noise reduction of 21.8 dB. The proposed algorithm does slightly better at low \acp{SNR} ($-10$ dB and lower).  At an \ac{SNR} of $-5$ dB, the fixed-\ac{MVDR} is slightly better and as the \ac{SNR} increases the proposed algorithm's noise reduction drops by several decibels (reaching 17.6 for an \ac{SNR} of 10 dB).  This is not decidedly troublesome since at high \acp{SNR} the issue of noise reduction is of lesser consequence.

For the case of moving interference, the proposed algorithm significantly outperforms fixed-\ac{MVDR}. The fixed-\ac{MVDR} algorithm reduces noise by 16.1 dB, whereas the proposed algorithm yields a reduction of 29.3 at $-20$ dB. As the \ac{SNR} increases, the noise reduction gradually decreases but typically remains higher than the fixed-\ac{MVDR}.  For example, at \acp{SNR} of $-10$, 0, and 10 dB the respective noise reductions are 26.2, 22, and 18.5 dB.  Due to the changing nature of the interference, the initial covariance estimate of the fixed-\ac{MVDR} algorithm is deficient.  In contrast, the proposed algorithm constantly adapts and consequently manages to effectively reduce noise.  We note that the proposed algorithm is more successful in this dynamic case than in the case of 3 static interferers.  This can be explained by the increased challenge of 
suppressing multiple sources.

The proposed algorithm significantly outperforms the fixed-\ac{MVDR} algorithm in the scenario of a moving interferer with respect to \ac{STOI}.  Interestingly, in the static scenario the two algorithms have virtually indistinguishable \ac{STOI} scores.  This is despite the fact that there are differences in their noise reduction.

\bed{We now discuss the \red{performance} of the two algorithms \red{tested} with a \ac{RA}.
	\red{These} are labeled `oracle adaptation (RA)' and `proposed algorithm (RA)'
	in Figs \ref{fig:resultsN},  \ref{fig:resultsD}, and \ref{fig:resultsS}.
	The noise reduction attainable with the reduced array with the proposed algorithm
	is roughly 6 dB which is close to the limit set by oracle adaptation with a reduced array.
	Full use of all channels form the \acp{AVS} provided an improvement of approximately
	15 to 25 dB.
	The performance with respect to \ac{STOI} with a reduced array is only slightly
	better than the unprocessed signal.
	The distortion levels of the reduced array are 
	in the vicinity of -30 dB which is an improvement over the full array with distortion of
	approximately -18 to -20 dB.  This improvement is apparently due to the 
	fact that the \emph{unprocessed signal} was defined as the average of the
	two omnidirectional channels used by the reduced array.
	In any case, the full array preforms satisfactorily in terms of distortion
	and improvement is of negligible significance;  utilization of all channels
	\emph{does} provide significant improvements with respect to noise reduction and \ac{STOI}. }

\subsection{Threshold parameter sensitivity}
In this subsection, we examine the impact of the threshold parameter $\eta$ on the performance of the proposed algorithm.  If $\eta$ is set too low, then too many bins are mistakenly labeled as containing desired speech.  This may lead to poor noise estimation since fewer bins are used in the estimation process.  Conversely, if $\eta$ is set too high, bins which do contain desired speech will not be detected as such which may lead to contamination of the noise estimation (as seen in the \ac{MPDR} based algorithms).  Presumably, a certain region of values 
in between these extremes
will yield desirable results with respect to the conflicting goals.

We repeatedly executed the algorithm with $\eta$ taking on different values (the other parameters in Table \ref{table:params} remain unchanged).  This was done for \acp{SNR} ranging from $-20$ dB to 10 dB.  The noise reduction results are plotted in Fig.~\ref{fig:paramN}, the distortion results in Fig.~\ref{fig:paramD}, and the STOI measure in Fig.~\ref{fig:paramS}.  The \ac{STOI} measures peak in the vicinity of $\eta = 0.9$ and the curve is fairly flat indicating robustness.  Similarly, $\eta = 0.9$ is a fairly good choice with respect to noise reduction, although for low \acp{SNR} Fig.~\ref{fig:paramN} indicates that a slight increase in $\eta$ is beneficial for low \acp{SNR} and conversely a slight decrease in $\eta$ is beneficial for high \acp{SNR}.  

The distortion levels are somewhat better in the vicinity of $\eta$ = 0.6.  However, since the distortion is minor, this slight improvement does not justify the accompanying degradation in noise and \ac{STOI} which are of notable quantity.

\begin{figure}[!h]{
		\subfigure[Noise reduction]{
			\includegraphics[width = 0.9\linewidth]{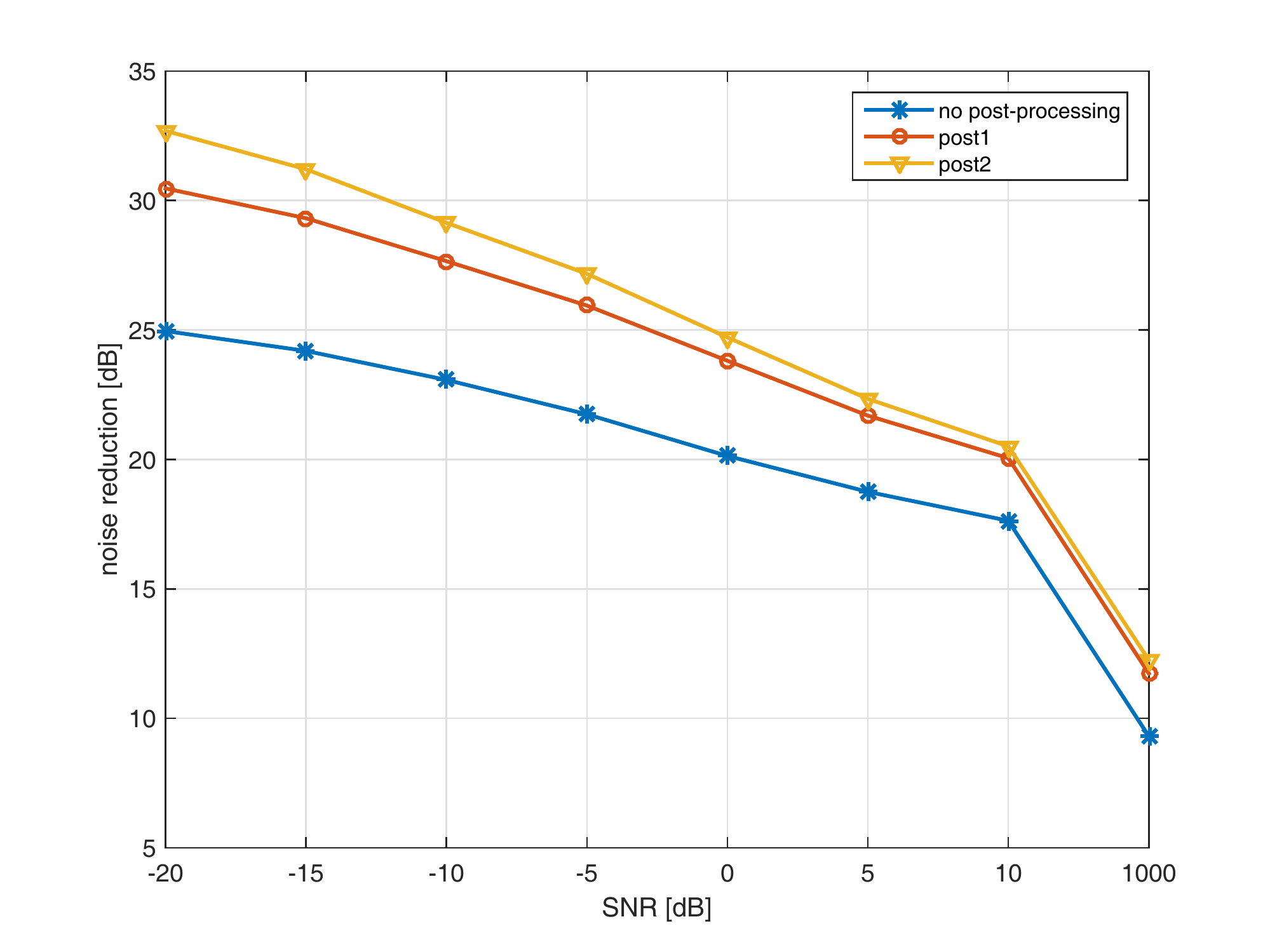} \label{Fig:ResultsA}%
		}\vspace{-2mm}
		\subfigure[Distortion]{
			\includegraphics[width = 0.9\linewidth]{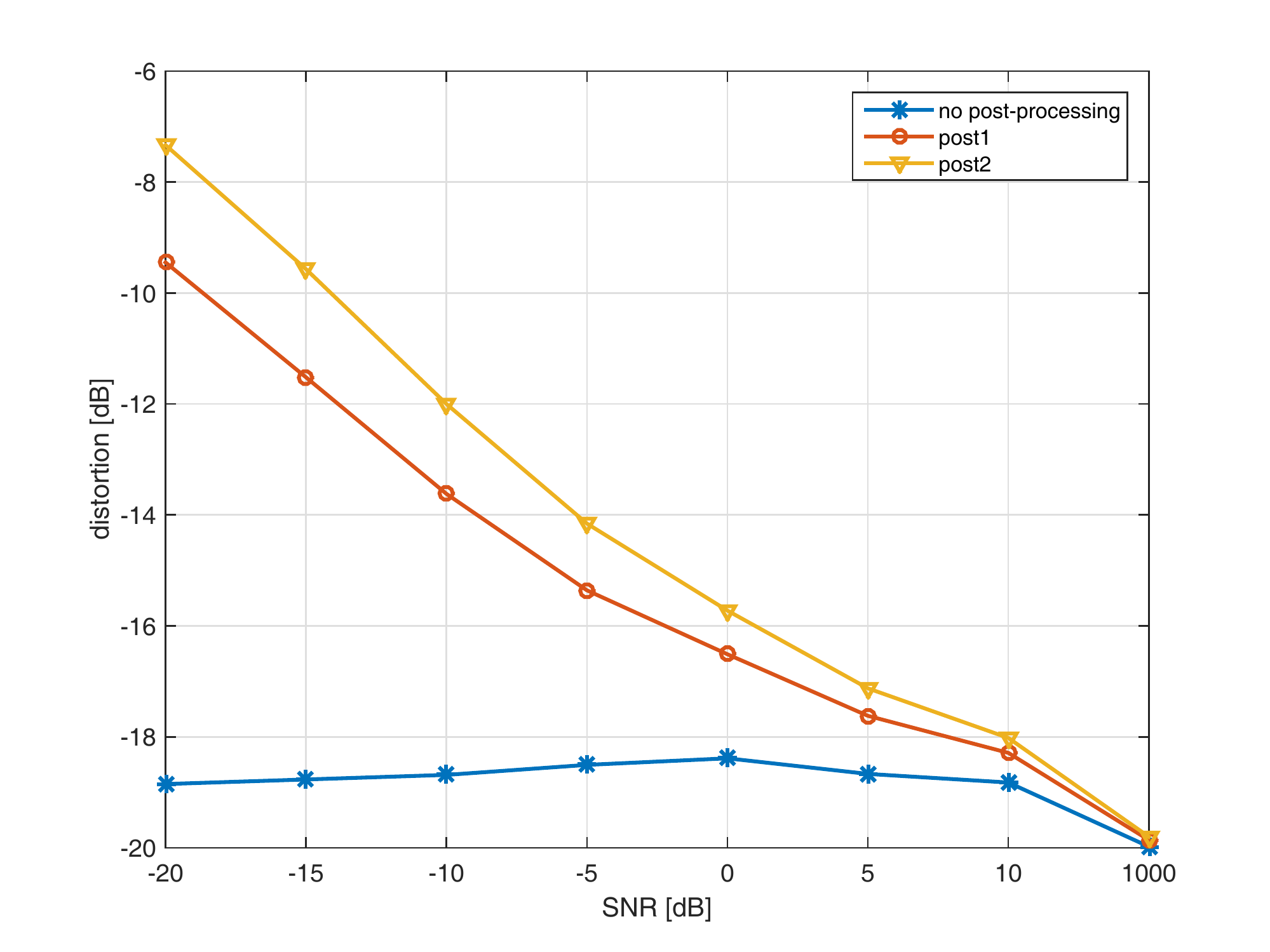} \label{Fig:ResultsB}%
		}\vspace{-2mm}
		\subfigure[STOI]{
			\includegraphics[width = 0.9\linewidth]{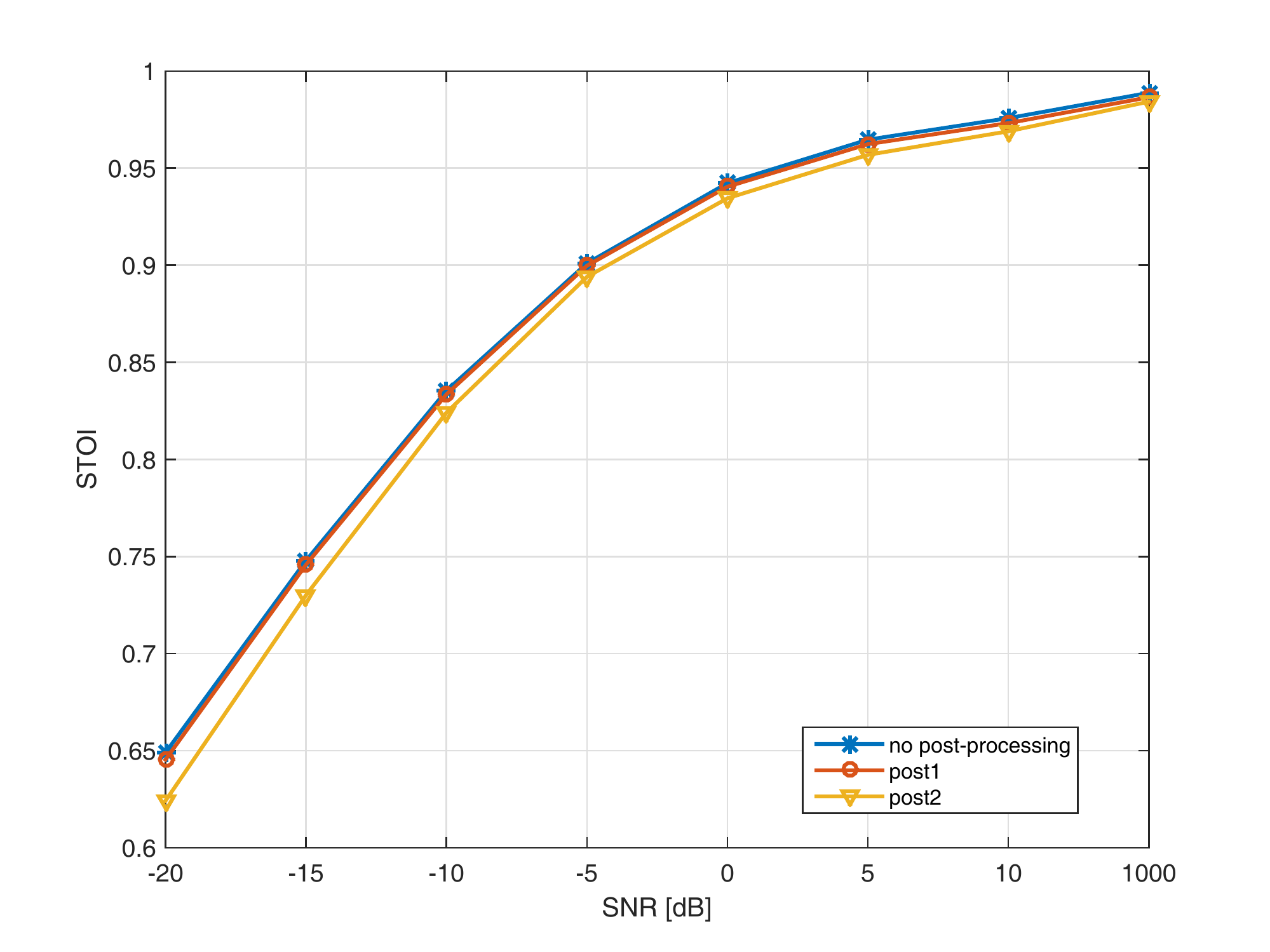} \label{Fig:ResultsC}%
		}\caption{Effects of post-processing on noise reduction, distortion, and \ac{STOI}.}\label{fig:post}}
\end{figure}

\subsection{Post-processing results}\label{sec:evalpost}
In this subsection, we examine the effects of post-processing.
Three parameters influence the post-processing: 
$\beta$, $\mathrm{SNR_{min}}$, and $\WF_{\mathrm{min}}$.
\bed{Setting the latter two parameters at lower values
	corresponds to a more aggressive suppression of noise,
	whereas higher values correspond to a more conservative 
	approach regarding signal distortion.  To illustrate this
	trade-off,}
   we test the two sets of parameters whose values\footnote{It
	should be noted that
	$\mathrm{SNR_{min}}$ describes the ratio of \emph{powers} whereas
	$\WF_{\mathrm{min}}$ describes a filter's \emph{amplitude}.  Consequently,
	the former is converted to decibel units via $10\log_{10}(\bdot)$, and
	the latter by  $20\log_{10}(\bdot)$.}
	 are given in
Table \ref{table:post}.  These two sets are referred to as `post$_1$' and
`post$_2$', respectively.
\bed{In general, post-processing parameters are determined empirically;
	the designer tests which values yield results which are satisfactory
	for a particular application.}
\begin{table}[!h]
	\centering
	\caption{Parameter values in post\red{-}processing.} \label{table:post}
	\rowcolors{1}{}{lightgray}
	\begin{tabular}{l l l}
		\textbf{Parameter:}    & $\bf{post_1}\textbf{:}$ &$\bf{post_2}\textbf{:}$\\ \hline
		$\beta$                & ~~0.9              &~~0.9 \\ 
		$\mathrm{SNR_{min}}$   & -10 dB             &-24 dB \\
		$\WF_{\mathrm{min}}$   & \phantom{1}-8 dB   &-20 dB \\
	\end{tabular}
\end{table}   

Figure \ref{fig:post} portrays the effects of post-processing (using the three speaker scenario as a test case) on the performance.  Post-processing reduces noise but increases distortion and adversely affects intelligibility as measured by \ac{STOI}
(this degradation is very minor for `post$_1$' and more prominent in `post$_2$').  
The parameters of `post$_1$' are more conservative and the parameters of `post$_2$'
are more aggressive with respect to noise reduction.  The former do not reduce as much noise, but have less distortion and only a minor degradation of \ac{STOI} score.
The latter reduce more noise at the expense of greater distortion and lower \ac{STOI}.  With the latter, audio artifacts have a stronger presence than the former. In general, the parameters may be adjusted to attain a desirable balance.   
  

\section{Conclusion} \label{sec:conclusion}
We proposed an array which consists of two \acp{AVS} mounted on an eyeglasses frame.
This array configuration provides high input \ac{SNR} and removes the need for tracking changes in the steering vector.  An algorithm for suppressing undesired components was also proposed.  This algorithm adapts to changes of the noise characteristics by continuously estimating the noise covariance matrix.  A speech detection scheme is used to identify the presence of time-frequency bins containing desired speech and preventing them from corrupting the estimation of the noise covariance matrix.
The speech detection plays a pivotal role in ensuring the quality of the output signal; in the absence of a speech detector, the higher levels of noise and distortion which
are typical of
\ac{MPDR} processing are present.
Experiments confirm that the proposed system performs well in both static and changing scenarios.  The proposed system may be used to improve the quality of 
speech acquisition in smartglasses.

\appendix

\bed{\section{\red{Background} on \acp{AVS}}}\label{sec:append}
\bed{A \red{sound field can be described as a combination} of two fields which are coupled:
a \emph{pressure} field and a \emph{particle-velocity} field.
The former is a scalar field and the latter is a 
vector field consisting of three Cartesian components.}

\bed{Conventional sensors which are typically used in
acoustic signal-processing measure the pressure field.
Acoustic vectors sensors (AVSs) also measure the particle-velocity field,
and thus provide more information:  each sensor provides
four components rather than one component.}

\begin{figure}
	\centering
	\includegraphics[trim={0 7mm 0 15mm},
	clip,width=0.7\linewidth]{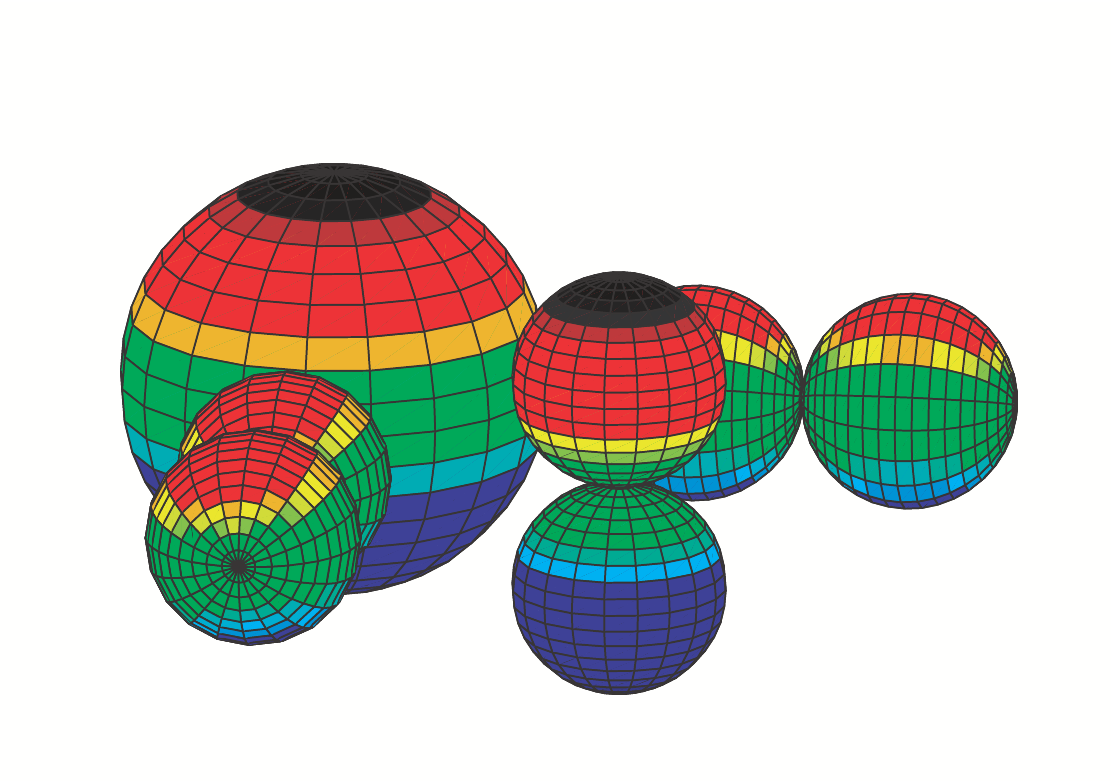}\label{BGC_fig:mondip}
	\caption{The magnitude of the directivity patterns of an \ac{AVS} are plotted. They consist of a monopole and three mutually orthogonal dipoles.}
	\label{fig:MonDip}
\end{figure}

\bed{An \ac{AVS} consists of four collocated subsensors: one monopole and three
	orthogonally oriented dipoles.  For a plane wave, each subsensor has a distinct
	directivity response.  The response of a monopole element is} 
	\begin{equation}
	D_{\mathrm{mon}} = 1\,,
	\end{equation}
	and the response of a dipole element is
	\begin{equation}
	 D_{\mathrm{dip}} = \bq^T \bu \,,
	\end{equation}
	\bed{where $\bu$ is a unit-vector denoting the wave's \ac{DOA}, and
	$\bq$ is a unit-vector denoting the subsensor's orientation.
	From the definition of scalar multiplication, it follows that
	 $D_{\mathrm{dip}}$ corresponds to the cosine of the angle between the
	 signal's \ac{DOA} and the subsensor's orientation.
	 The orientation of the three subsensors are
	 $\bq_x = [1\;0\;0]^T$, $\bq_y = [0\;1\;0]^T$, and $\bq_z = [0\;0\;1]^T$.
	 The monopole response, which is independent of \ac{DOA},
	 corresponds to the pressure field and the three dipole responses
	 correspond to a \emph{scaled} version of the Cartesian particle-velocity components.
	 Fig.~\ref{fig:MonDip} portrays the magnitude of the four spatial responses.} 

\bed{For a spherical wave, the acoustical impedance is frequency-dependent.
It can be shown that the dipole elements undergo a relative gain of 
 $1 +\frac{c}{r}\frac{1}{j\omega} $ over an omnidirectional sensor
 (as discussed in Sec.~\ref{sec:array}).  This phenomenon is 
 manifest\red{ed particularly} at lower frequencies for which
 the wavelength is significantly shorter
 tha\red{n} the source-receiver distance.}

\bed{A standard omnidirectional microphone functions as a monopole element. 
	Several approaches are available for constructing the dipole components of an \ac{AVS}.
	One approach applies differential processing of
	closely-spaced omnidirectional sensors \cite{Olson46, Elko95, Elko97, Derkx09}. 
	An alternative approach employs acoustical sensors with inherent directional
	properties \cite{Shaju09, Derkx10}.
    Recently, an \ac{AVS} based on \ac{MEMS} technology has been developed \cite{deBree03} 
    and \red{has} become commercially available.  The experiments discussed in Sec.~\ref{sec:evaluation}
    use such devices.}
   
\bed{The different approaches mentioned produce approximations of the ideal responses.
	For instance, the subsensors can be placed close to each other but are not strictly
	collocated; spatial derivatives are estimated, etc.
	The approaches mentioned above differ  with respect to attributes such as robustness, sensor noise, and cost.  A discussion of these characteristics is beyond the scope of the current paper.}


\vspace{-3.5mm}
\section*{Acknowledgments}
\vspace{-1.5mm}
We wish to acknowledge technical assistance from Microflown Technologies (Arnhem, Netherlands) related to calibration and reduction of sensor noise.  This contribution was essential for attaining quality audio results.
\vspace{-3.5mm}
\FloatBarrier
\balance
	  \bibliographystyle{elsarticle-num} 
	  \bibliography{abib}
	
	
	%
	%
	%
\end{document}